\documentclass[sigconf]{acmart}

\AtBeginDocument{%
  }

\setcopyright{acmlicensed}
\copyrightyear{2025}
\acmYear{2025}

\acmDOI{XXXXXXX.XXXXXXX}

\acmConference[SIGMOD'25]{the International Conference on Management of Data}{June 22-27, 2025}{Berlin, Germany}

\usepackage{ulem} 
\usepackage{amsmath,amsfonts}
\usepackage[ruled,linesnumbered]{algorithm2e}
\usepackage{algorithmic}
\usepackage{graphicx}
\usepackage{textcomp}
\usepackage{xcolor}
\usepackage{url}
\usepackage{ragged2e,caption}
\usepackage{bbding}
\usepackage{amsmath}
\usepackage{textcase}
\usepackage{enumitem}

\usepackage[subtle]{savetrees}

\usepackage{booktabs} 
\usepackage{times,epsfig}
\usepackage{enumerate}
\usepackage{subfigure}
\usepackage{multirow}
\usepackage{tabularx, subfigure}
\usepackage{hyperref}
\usepackage{float} 
\usepackage{array}
\usepackage{makecell}
\newcommand{\centered}[1]{\begin{tabular}{c} #1 \end{tabular}}

\usepackage{color}

\newcommand{\eat}[1]{}

\PassOptionsToPackage{normalem}{ulem}
\usepackage{ulem}

\renewcommand{\emph}[1]{\textit{#1}}

\newcommand{\sysname}{PoneglyphDB}

\begin{document}
\title{{\sysname}: Efficient Non-interactive Zero-Knowledge Proofs for Arbitrary SQL-Query Verification}

\author{Binbin Gu}
\affiliation{%
  \institution{Unversity of California, Irvine}
  \streetaddress{}
  \city{Irvine}
  \country{USA}
  \postcode{}
}
\email{binbing@uci.edu}

\author{Juncheng Fang}
\orcid{}
\affiliation{%
  \institution{Unversity of California, Irvine}
  \streetaddress{}
  \city{Irvine}
  \country{USA}
}
\email{junchf1@uci.edu}

\author{Faisal Nawab}
\orcid{}
\affiliation{%
  \institution{Unversity of California, Irvine}
 \city{Irvine}
  \country{USA}
}
\email{nawabf@uci.edu}

\renewcommand{\shortauthors}{Trovato and Tobin, et al.}

\begin{abstract}
In database applications involving sensitive data, the dual imperatives of data confidentiality and provable (verifiable) query processing are important.  This paper introduces \sysname, a database system that leverages non-interactive zero-knowledge proofs (ZKP) to support both confidentiality and provability. Unlike traditional databases, \sysname~ enhances confidentiality by ensuring that raw data remains exclusively with the host, while also enabling verifying the correctness of query responses by providing proofs to clients. 

The main innovation in this paper is proposing efficient ZKP designs (called circuits) for basic operations in SQL query processing. 
These basic operation circuits are then combined to form ZKP circuits for larger, more complex queries. {\sysname}'s circuits are carefully designed to be efficient by utilizing advances in cryptography such as PLONKish-based circuits, recursive proof composition techniques, and designing with low-order polynomial constraints.
We demonstrate the performance of \sysname~ with the standard TPC-H benchmark. Our experimental results show that \sysname~ can efficiently achieve both confidentiality and provability, outperforming existing state-of-the-art ZKP methods. 
\end{abstract}


\maketitle

\vspace{-0.1in}
\section{Introduction}
As databases serve as the backbone of diverse applications, the responsibility entrusted to them extends beyond data storage---it encompasses the safeguarding of sensitive information. This is an important consideration, especially for databases that store personal data. Also, in applications with sensitive information such as census data or unemployment statistics, 
it is important for users to trust that the database owner is using the correct database (i.e., one that is consistent with the claimed database) and that the provided answers accurately reflect the intended computations.



We tackle the problem of providing a database solution to ensure two characteristics: \textbf{(1)~Confidentiality}: this property means that raw data is only maintained at the host (i.e., service provider) and is not shared or made public to any other node. Other nodes only receive responses to specific queries that they send to the host node and that the host node agrees to process. \textbf{(2)~Provability:} ensuring the correctness of processing. This means that the results computed by the host node and sent back to users reflect the correct processing of the query on the private database. This requires that the host node provides a proof of the correctness of the query response to the user.


As an example, consider the healthcare sector, where medical research institutions collect patient data for collaborative studies. Institution X may wish to share insights about the data with data consumers Y, Z, and W without disclosing the raw data.
Data consumers send queries to X, where queries are processed and their responses sent back to the consumer. X has control on what queries to answer and therefore can control the confidentiality of data~\cite{byun2008purpose}. 
However, because data consumers do not have access to the raw database, they cannot verify the correctness of processing. 


\begin{sloppypar}
In this paper, we build on cryptographic solutions, namely \emph{zero-knowledge proofs (ZKP)}~\cite{ZK-origins, fiege1987zero}, to address the challenges of ensuring both confidentiality and provability in database query processing~\cite{derei2023scaling, yue2023veribench, yue2022glassdb}. ZKP constitutes a powerful cryptographic tool that enables one party, the prover, to convince another party, the verifier, of the correctness of a statement without revealing any specific information about the statement itself. 
%
%
The protocols used in ZKP systems can be made interactive and non-interactive ZKP. Interactive ZKP utlize cryptographic protocols wherein two entities, namely a \emph{prover} and a \emph{verifier}, dynamically exchange messages to establish the validity of a statement without revealing any sensitive information. Previous research has introduced interactive ZKP for verifying SQL queries~\cite{zhang2015integridb, zhang2017vsql, li2023zksql}. 
The rationale behind using interactive ZKP for SQL query verification (like ZKSQL~\cite{li2023zksql}) is that they enable the prover to engage with the verifier in multiple rounds, incrementally constructing and verifying parts of the proof. This step-by-step interaction often results in smaller circuit sizes, as the prover can break down the computation into manageable parts rather than generating a single large, complex proof upfront. However, the interactive nature introduces challenges related to synchronization and availability between the prover and verifier.
Specifically, interactive ZKP are accompanied by two main drawbacks that currently limit their widespread adoption in practical applications: (1)~They necessitate interaction between the prover and verifier throughout the proof protocol, imposing requirements for availability and synchronization between the involved parties. This challenge exacerbates failures and timeouts. (2)~They lack transferability, as only the specific verifier(s) engaged in the original protocol possess the capability to verify the proof. This excludes the possibility of reusing (and caching) previously computed responses and proofs for multiple verifiers.
\end{sloppypar}


On the other hand, non-interactive ZKP is a cryptographic protocol class designed to establish the validity of a statement without the necessity for a dynamic exchange of messages between the prover and verifier. Unlike interactive ZKP, non-interactive ZKP empowers a prover to generate a single, self-contained proof that can be subsequently verified by any party possessing the necessary verification key. 
This characteristic eliminates the requirement for real-time interaction during the proof protocol, providing greater flexibility and efficiency in scenarios where asynchronous verification or limited interactivity is desired. 
Non-interactive ZKP, however, can experience significant overheads 
if the underlying cryptographic circuits are not carefully optimized.


%


In this paper, we propose a database system named \sysname, which incorporates non-interactive ZKP and the recursive proof composition technique. \sysname~ achieves both confidentiality and provability through non-interactive ZKP. 
%
%
%
{\sysname} differs from traditional databases by incorporating ZKP \emph{circuits}, which are sets of equality constraints over arithmetic expressions designed to mimic the steps of query processing. These circuits enable ZKP frameworks to generate proofs of correctness for computations. In \sysname, we design circuits to represent basic query operations such as range checks, sorting, group-by, and joins, which are combined to handle more complex queries.
Circuit design directly impacts proof generation performance. We optimize \sysname's circuits using the PLONKish framework~\cite{plonkish}, considering factors such as circuit size, polynomial degree, and batching construction. 
Additionally, {\sysname} leverages a recursive structure for composing proofs of multiple statements, reducing the overall proof size and computational overhead. This is made possible by advancements in proving systems that utilize recursive proof composition techniques~\cite{bowe2019recursive, kothapalli2022nova}.

Table~\ref{table:method_comparison} compares \sysname~and prior research on verifiable database systems~\cite{zhang2015integridb, zhang2017vsql, li2023zksql} in terms of three properties: (1) the assurance of zero-knowledge (i.e., confidentiality), (2) non-interactive operability, and (3) applicability to arbitrary SQL queries.
The protocols used in vSQL and vSQL+ are originally presented in an interactive manner, however they can be made non-interactive via the Fiat-Shamir heuristic~\cite{fiat1986prove}. 
To the best of our knowledge, {\sysname} is the first system that achieves all these desirable properties. 
Also, in the evaluation section, we show that {\sysname} performance outperforms the state-of-the-art ZKP methods.




This paper is structured as follows: Section~\ref{sec:background} presents background material. Section~\ref{sec:workflow} proposes the general design of {\sysname} followed by the detailed design in Section~\ref{sec:gates}.
Section~\ref{sec:evaluation} presents our experimental evaluations. Section~\ref{sec:related_work} presents related work and Section~\ref{sec:conclusion} concludes the paper.


\begin{table}
\centering
\resizebox{0.9\columnwidth}{!}{
\begin{tabular}{|c|c|c|c|}
\hline
\bfseries  & \bfseries \makecell[c]{Zero-knowledge} & \bfseries \makecell[c]{Non-interactive}  & \bfseries \makecell[c]{Arbitrary SQL queries}  \\\hline

\makecell{\bfseries IntegriDB~\cite{zhang2015integridb}}  & \XSolidBrush  & \XSolidBrush & \XSolidBrush  \\\hline

\makecell{\bfseries vSQL~\cite{zhang2017vsql} }  & \XSolidBrush  & \XSolidBrush & \Checkmark         \\\hline

\makecell{\bfseries vSQL+~\cite{zhang2017zero} }  & \Checkmark  & \XSolidBrush & N/A         \\\hline

\makecell{\bfseries ZKSQL~\cite{li2023zksql} }  & \Checkmark  & \XSolidBrush & \Checkmark  \\\hline

\makecell{\bfseries \sysname}  & \Checkmark  & \Checkmark & \Checkmark       \\\hline
\end{tabular}
}

\caption{ Comparing cryptography-based methods for verifiable SQL queries.}
\vspace{-0.4in}
\captionsetup{skip=2pt} 
\label{table:method_comparison}
\end{table}

\vspace{-0.1in}
\section{Preliminaries}
\label{sec:background}

\subsection{Zero-Knowledge Proofs}
In the area of ZKP, a prover can convincingly demonstrate the truth of a given statement to a verifier without divulging any additional information. Specifically, ZKP empowers a prover $\mathcal{P}$ with a private witness $\mathnormal{w}$ (a ``private witness" refers to a piece of secret information such as a secret input, e.g., database) to validate the truth of a public statement $\mathcal{F}$ (with respect to $\mathnormal{w}$) to a verifier $\mathcal{V}$, while preserving the confidentiality of the underlying information $w$. 
For instance, suppose a database contains information about employees, and the data owner wants to prove to the verifier that the average salary of the employees is a certain number without disclosing individual salary details. In this scenario, the private witness $\mathnormal{w}$ is the information about individual salaries, and the public statement $\mathcal{F}$ is the average salary computed by the SQL query.

In scenarios with multiple verifiers, such as healthcare settings where various institutions need to verify query results, interactive protocols can become impractical due to the need for multiple rounds of communication with each verifier. Non-interactive ZKP addresses this issue by allowing the prover to generate a single proof that can be verified by all parties without further interaction. For public-coin interactive protocols~\cite{ZK-origins}, the Fiat-Shamir heuristic~\cite{fiat1986prove} effectively converts them into non-interactive ones (e.g., vSQL), maintaining efficiency and minimizing computational overhead while supporting asynchronous verification.

\subsection{Arithmetization}
\label{sub:plonkish}
The use of arithmetic circuits is the most common paradigm for expressing computations within ZKP systems. 
We introduce the PLONKish arithmetization~\cite{plonkish} that we use in \sysname. We emphasize that the PLONKish circuits used in PoneglyphDB serve
primarily as a vehicle to demonstrate the potential of non-interactive
ZK proofs within database management systems (DBMSs), rather
than representing a state-of-the-art protocol.

PLONKish circuits are defined in terms of a rectangular matrix of values. We refer to rows, columns, and cells of this matrix with the conventional meanings.
In the following, we introduce the key definitions of PLONKish circuits that are relevant to our circuit optimization. 

\begin{itemize}[leftmargin=*]
    \item [1.] {\bf Fixed columns.} Fixed columns are fixed by the circuit. The values in the fixed columns are usually constants. 

    \item [2.] {\bf Advice columns.} Advice columns correspond to \emph{witness} values. These are private inputs and intermediate values generated during the circuit computation.

    \item [3.] {\bf Instance columns.} Instance columns are used for any elements shared between the prover and verifier. In most cases, they are used for public inputs and outputs. 

    \item [4.] {\bf Equality constraints.} Equality constraints specify that two given cells must have equal values. 

    \item [5.] {\bf Polynomial constraints.} For each row in the matrix, the multivariate polynomials over the field $\mathbb{F}$ must be evaluated to zero. 
    
 
\end{itemize}

\begin{figure}[t]
\centering
\includegraphics[width=0.5\textwidth]{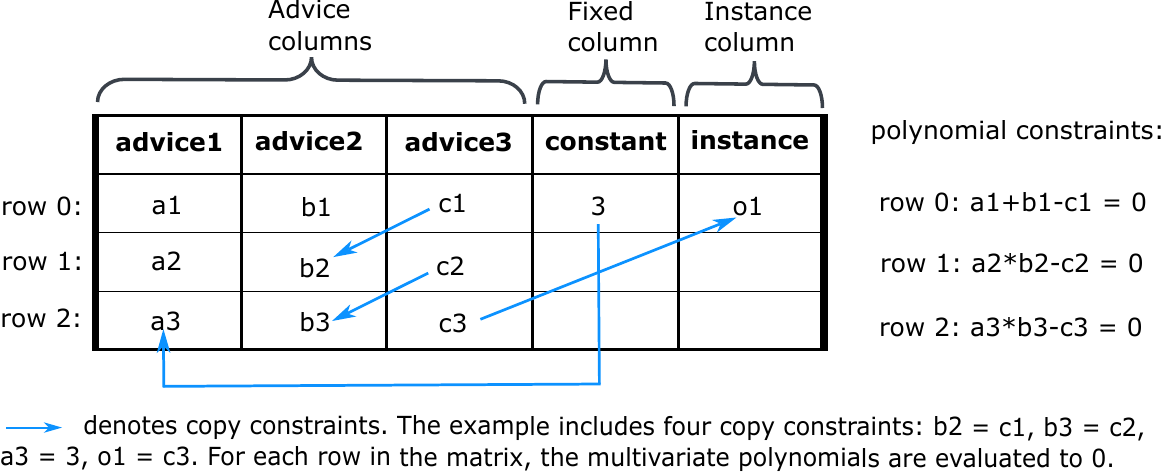}
\vspace{-0.1in}
\captionsetup{skip=2pt} 
\caption{PLONKish circuits illustration for calculating the function $f(x, y, z) = 3*(x+y)*z$.}
\vspace{-0.2in}
\label{fig:plonkish}
\end{figure}

\begin{example}
Figure~\ref{fig:plonkish} illustrates the PLONKish circuit for calculating the function $f(x, y, z) = 3 * (x + y) * z$. The circuit utilizes three advice columns---advice1, advice2, and advice3---to store the private inputs and intermediate values during computation.

In row 0, the values for $x$ and $y$ are put into the cells in advice1 and advice2, respectively (i.e. $x = a1$ and $y = b1$). A polynomial constraint is introduced to ensure that $c1 = a1 + b1$, calculating $x + y$.
Next in row 1, the value of the input $z$ (i.e. $z=a2$) is put into the cell in advice1. Since $a1 + b1$ was computed as $c1$, its value is propagated to $b2$ and an equality constraint sets $b2 = c1$. A multiplier gate (with polynomial constraints) then calculates $c2 = a2 * b2$, evaluating $a2*b2-c2 = 0$.
Finally in row 2, the constant $3$ is brought into advice1 at the cell of $a3$ and fixed by the copy (or equality) constraint $a3 = 3$. The previous intermediate result $(a1 + b1)*a2 = c2$ is copied into $b3$, with an equality constraint $b3 = c2$. The final output cell $c3$ calculates $a3 * b3$, which evaluates to $3 * (a1 + b1) * a2$.

This result is copied into the public instance column at $o1$, allowing the verifier to read off the final output. 
Through a series of advice columns, equality constraints, and polynomial operations, the circuit computes the desired function in a modular fashion while keeping intermediate values private. The verifier only sees the final outputs revealed in the instance column.
\end{example}

\vspace{-0.1in}
%
To facilitate the modularity of PLONKish circuits, the designs of basic functions are represented as \emph{gates}. A gate is a collection of columns and constraints that together implement a basic operation, such as division and multiplication. In our case, we will define gates for basic operations for query processing such as range checks and sorting. Gates can then be combined to implement more complex functions (or more complex gates).

\vspace{-0.1in}
\section{System Overview}

\label{sec:workflow}

\subsection{System Model}
In \sysname, 
the prover $\mathcal{P}$ hosts a copy of a private database $\mathcal{DB}$. The prover receives queries from one or more verifiers, collectively denoted $\mathcal{V}$. The prover does not provide raw access to the data in the database. However, it answers queries sent by the verifiers.

\subsection{Workflow Overview}

\begin{figure}[t]
\centerline{\psfig{figure=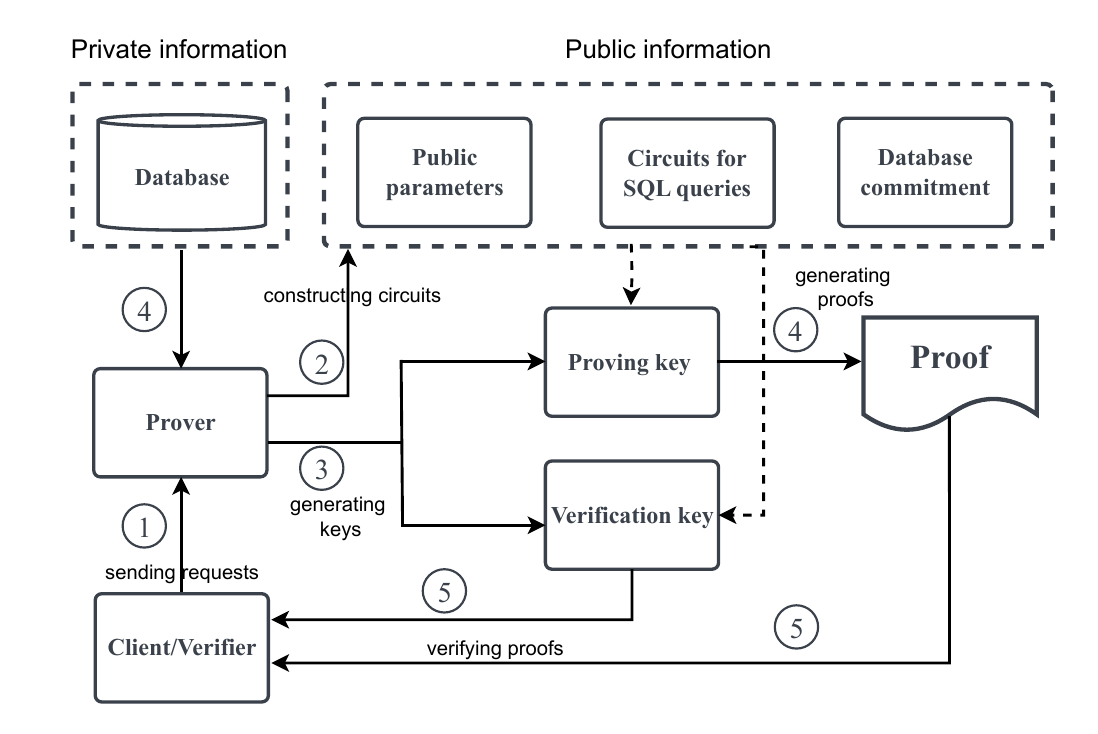, width=0.42\textwidth} }
\vspace{-0.2in}
\caption{\sysname~ Workflow.}
\label{fig:workflow}
\vspace{-0.3in}
\end{figure}

\sysname~ operates in five key phases, as illustrated in Figure \ref{fig:workflow}:

\textbf{(1) Sending SQL Queries:} The client, who will eventually assume the role of a verifier, sends SQL queries for execution against a private database, directly to the prover. The prover, holding exclusive access to this private database, is tasked with executing the queries.

{\bf (2) Circuit Construction:} 
Upon receipt of the query request, the prover is then responsible for constructing the SQL queries and database commitment into arithmetic circuits. 
These circuits, delineated by gates and polynomial constraints, performs the computations that need to be proven.
%
%
The SQL circuit encodes the desired SQL logic to be evaluated. This circuit allows for the computation of different inputs provided by the verifiers during the verification phase.

A database commitment is a cryptographic representation of a database state, enabling proof of properties about the data without revealing the data itself. This allows the prover to include evidence in the proof that the query was indeed processed on $\mathcal{DB}$. We employ the Inner Product Argument (IPA) protocol~\cite{bootle2016efficient}, operating over a 254-bit prime field, to generate this commitment. 
We choose the IPA protocol for the following reasons: (1) the proving time is typically linear with respect to the circuit size or the degree of the committed polynomial, (2) the proof size and verification time are logarithmic in the circuit size, due to the recursive structure of the inner product proof~\cite{bowe2019recursive}, and (3) it is compatible with PLONKish-based circuit designs. 

By leveraging IPA, we can encode both the database commitment and the desired SQL logic within a unified framework~\cite{bootle2016efficient, bunz2018bulletproofs}.
%
%
The public parameters are a shared foundation for both provers and verifiers in proof creation and validation. This process utilizes publicly verifiable randomness and avoids the need for a trusted setup~\cite{chiesa2020marlin}. Initial proof creation uses only public information, forming the basis for subsequent proofs~\cite{bowe2019recursive}.
%



{\bf (3) Key Generation:} 
Leveraging the public parameters and the circuit description, \sysname~ generates a proving key. This key is used to generate proofs corresponding to the circuit. Concurrently, a verification key is also produced, grounded in the same public parameters and circuit description. This verification key empowers verifiers to authenticate the proofs associated with this circuit. 

{\bf (4) Proof Generation:} 
The prover uses the proving key to generate a ZKP validating the correct computation of the SQL query over the private database. Leveraging the previously established database commitment, the prover creates a verifiable link between the committed database, query execution, and result. This process incorporates commitments to relevant database rows and intermediate computation steps, forming a chain of verifiable commitments from the initial database state to the final query output. 

{\bf (5) Proof Verification:} Finally, the verifier employs the public verification key to efficiently validate the ZKP received from the prover. The verification algorithm essentially interpolates the wiring polynomials and checks that all constraints are satisfied.
%

\textbf{{\sysname} Positioning in the Workflow.} 
In {\sysname}, we utilize existing work from the cryptography community to perform the steps of key and proof generation and verification. The main contribution of our work is in the design of SQL circuits that allow for efficient proving and verification given the utilized framework~\cite{halo2}.




\subsection{Application Framework Disuccsion}
\label{subsec:application}

\textbf{Trust Model.} In the PoneglyphDB framework, the core parties involved are the database owner, acting as the prover, and the clients, acting as the verifiers. The prover and verifiers do not trust each other as we detail below. There is an additional entity, the auditor, that both the prover and verifier trust (there can be multiple auditors). The following are trust concerns between the parties:
\begin{itemize}[leftmargin=*]
    \item [1.] {\bf The prover may use a fake database:} The client does not trust the prover to be using the correct database to run the query. For example, a server responsible for running SQL queries might use a tampered or fake database to process a query and provide incorrect results to the verifier.

    \item [2.] {\bf The prover may process the query incorrectly:} Even if the correct database is used, the client cannot trust the accuracy of the results returned by the prover. The server could mistakenly or intentionally alter or fabricate the query results.

    \item [3.] {\bf The verifier may attempt to leak data:} On the other hand, the prover fears that the client might attempt to extract or infer additional information about the underlying database beyond what is allowed by the query. The client could craft a series of queries to uncover sensitive patterns, private data, or proprietary information, breaching the prover's confidentiality.  

\end{itemize}

We now map this trust model to a real-world scenario. In general, this trust model applies to cases when a database owner has sensitive data and wants to enable other entities (clients) to query the database without revealing data beyond what the database owner allows. In the healthcare scenario, a hospital H wants to provide query access to its database of patient data. It does not want to reveal all data, but would allow answering specific types of queries. The clients can be healthcare providers or research institutions that want to utilize H’s database for their research. In this scenario, an auditor can be a government or regulatory entity that both H and clients trust. Next, we map the trust model assumptions onto this example:

\begin{itemize}[leftmargin=*]
    \item [1.] The clients do not trust the hospital H to be using a correct database of patient’s data: The hospital H does not reveal the raw database to the clients, and therefore clients cannot attest to the authenticity of the database used to answer queries.

    \item [2.] The clients do not trust the hospital H to process queries on the database correctly: The hospital H may process the queries on the database incorrectly (e.g., by using approximation techniques to save costs, or by completely fabricating results).

    \item [3.] The clients may attempt to leak data from query responses: The clients may want to know more information about the database of H than what is provided in the answer to the query.

\end{itemize}

To address these trust issues, PoneglyphDB introduces the following measures:

\textbf{(1) Cryptographic Commitment to the Database:} PoneglyphDB requires the prover to make an irrevocable cryptographic commitment to the database. This ensures that the prover cannot substitute the real database with a bogus one while still producing valid query results. The commitment is made public and shared in an irrevocable and immutable manner, e.g., by utilizing a decentralized blockchain such as Ethereum. This ensures that the prover will use the same database that corresponds to the database commitment as clients have access to an irrevocable, immutable database commitment that they can compare with the commitment used in the received proof. The database commitment can also be audited by a third-party auditor (e.g., a regulatory or government entity trusted by both the prover and verifiers). The auditor in this case reads the raw database from the database owner, verifies its authenticity, and validates that the database commitment of the authentic database corresponds to the commitment that is shared in the blockchain and accessible by the verifiers. 

To prove the correctness of database updates and generate new commitments, a naive method would be to recompute the commitment for the entire database after each update. However, this approach is inefficient for large databases. A more advanced method might involve using a Merkle tree structure~\cite{DBLP:conf/sp/Merkle80}, where only the affected subtree is updated and recomputed, enabling efficient localized proof generation. Another potential method involves leveraging batch update techniques, where multiple updates are aggregated into a single proof.

\textbf{(2) Ensuring Query Result Correctness:} PoneglyphDB employs a constraint system that encodes SQL queries as circuits, ensuring that the verifier can confirm the query results are derived from a predefined and correct computation process. This process ensures the prover cannot return fabricated results.

\textbf{(3) Preserving Data Privacy:} To ensure the owner’s database privacy, PoneglyphDB employs ZKP. These proofs guarantee that the client only receives the query result without extracting or inferring any additional information about the underlying database. The zero-knowledge property protects the database from unwanted data leakage beyond what is revealed from the query response. This zero-knowledge property can be combined with other techniques that prevent other types of leakage of private data. For this reason, it is important to distinguish between the types of leakage that are prevented or allowed by the zero-knowledge property and complement it with other techniques. A ZKP does not reveal information beyond the query response, but it does reveal what is part of the query response (and anything that can be implied by the query response). Therefore, if there is data that a user should not have access to but is part of the query response, then it is leaked to the client. 
To prevent this type of leakage, ZKP can be combined with techniques such as access control, data masking, query filtering, and policy management to prevent processing queries on data that the user would not have access to ~\cite{agrawal2002hippocratic}. 
These techniques, typically applied as pre-processing steps, allow the prover to decide whether a query should be processed.
For example, if the client’s query asks to get raw data that should not be revealed, then the prover would not process the query.
Query filtering, in particular, helps prevent leakage by modifying the query plan to exclude sensitive information before execution, ensuring that only authorized data is included in the response.
Another type of leakage is to infer information about the database that were not intended to be revealed directly in the query response. This type of leakage can be prevented by incorporating differential privacy techniques as we discuss at the end of  Section~\ref{subsec:security}.

With these techniques, and reflecting back to the mapping to the real-world scenario above, we observe the following: (1) PoneglyphDB ensures that the hospital H uses an authentic database to process queries. This is by utilizing a trusted third-party auditor that verifies that the publicly shared irrevocable and immutable database commitment corresponds to an authentic database. If a prover attempts to utilize a different database, then the client is able to discover this as the client would match the publicly shared database commitment that is authenticated by an auditor with the database commitment that is used to process the query and received as part of the response. (2) PoneglyphDB ensures that hospital H processes the query correctly (according to the logic of the SQL query) by the ZKP constructions. (3) PoneglyphDB ensures that no information leakage occurs beyond what is revealed by the responses of hospital H. Hospital H can integrate further techniques---such as differential privacy---to ensure that other types of data leakage would not occur (as we describe in  Section~\ref{subsec:security}).


\subsection{Security Model}
\label{subsec:security}
\sysname~ leverages the Halo2 proving system~\cite{halo2}, which incorporates well-established cryptographic properties such as completeness, soundness, knowledge soundness, and zero-knowledge.

\begin{itemize}[leftmargin=*]
    \item {\bf Completeness.} If the prover can generate the PLONKish circuit (including the output) of a query, it can always convince the verifier that the PLONKish circuit of the query is true.

    \item {\bf Soundness.} For any false PLONKish circuit (including any wrong witness, inputs and output), the probability of a dishonest prover successfully convincing an honest verifier is negligible. 

    \item {\bf Knowledge Soundness.} When the verifier is convinced the PLONKish circuit is correct, the prover actually possesses a valid witness.

    \item {\bf Zero Knowledge.} The verifier learns only the information that can be inferred from the structure of the PLONKish circuit and the output of the query. No additional knowledge about the private witnesses or the database is revealed.

\end{itemize}

\begin{figure}[t]
\centerline{\psfig{figure=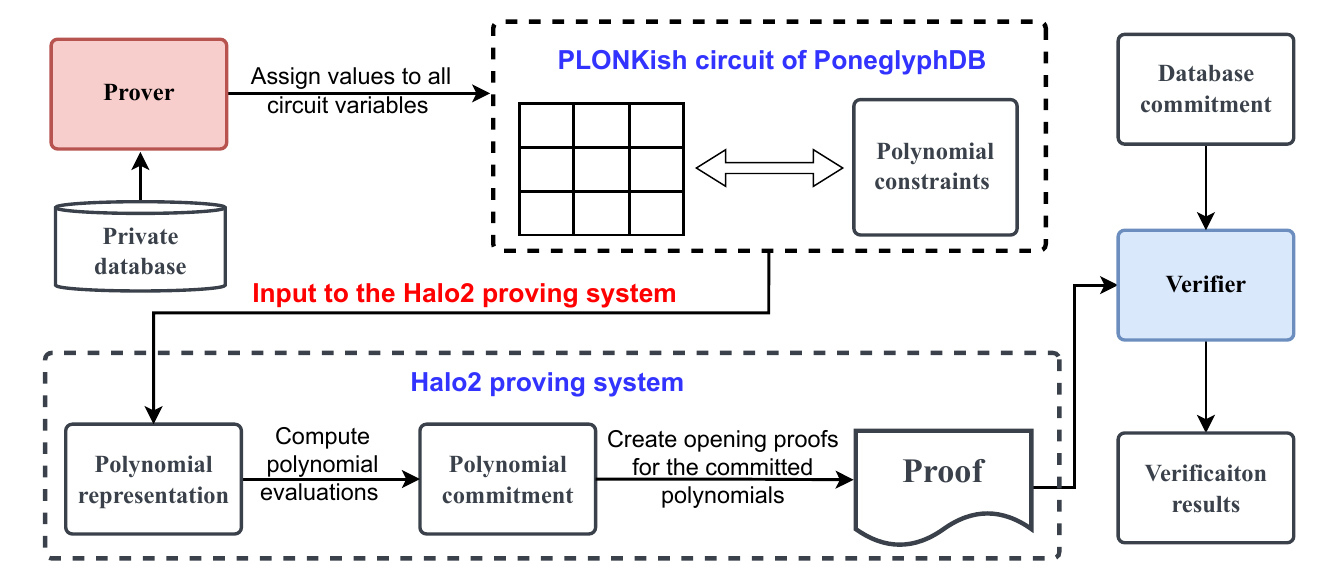, width=0.5\textwidth} }
\vspace{-0.2in}
\caption{Detailed components for generating ZK proofs.}
\label{fig:system}
\vspace{-0.2in}
\end{figure}

Figure~\ref{fig:system} illustrates the detailed components involved in generating ZKP within \sysname. The system comprises two main parts: the PLONKish circuit and the Halo2 proving system. The PLONKish circuit serves as the input to the Halo2 proving system, which then generates ZKP for the circuit. Our primary contribution lies in the design of the PLONKish circuits, which are tailored to optimize the performance and efficiency of the ZKP generation process.

%
The PLONKish circuit represents the computation 
%
%
of a SQL query. This circuit is a mathematical representation of the logical operations and constraints involved in the SQL query execution. The prover, who possesses the private database, assigns values to all circuit variables based on the actual data from the database. This step involves mapping the data inputs to the corresponding variables in the circuit, ensuring that the computation is correctly set up for proof generation. The prover must use the database agreed upon with the verifier by utilizing the previously established database commitment.

The {\it Polynomial representation} component, provided by the Halo2 proving system, translates the circuit into a polynomial form. This component encodes the computation and its constraints as polynomial equations, making them suitable for ZKP.

The {\it Polynomial commitment} component, also provided by the Halo2 proving system, allows the prover to commit to the polynomial evaluations without revealing the actual polynomials. This ensures that the prover cannot alter the polynomials after the commitment, maintaining the integrity of the proof.

The {\it Halo2 proving system} takes the committed polynomials and generates opening proofs. These proofs are designed to show that the committed polynomials satisfy the polynomial constraints derived from the PLONKish circuit.

It is important to note that the PLONKish circuits implemented in \sysname~ are primarily intended to illustrate the feasibility of using non-interactive ZK proofs within  DBMSs, rather than to claim they represent the latest advancements in ZK protocol design.

\textbf{Guarantees.} 
%
The correctness and security of \sysname~ depend on PLONKish circuits and the Halo2 proving system. we demonstrate the correctness of these circuits for SQL queries in Section~\ref{sec:gates}.

Regarding the security of the Halo2 proving system used in \sysname~, the prover and verifier engage in a non-interactive ZKP protocol utilizing Halo2's polynomial commitment scheme and recursive proof composition. 
The detailed security analysis of this protocol has been rigorously established in ~\cite{halo2-protocol, bowe2019recursive}. We present a high-level summary of the correctness of the system below.
To analyze the cryptographic protocols employed in Halo2, we utilize the Algebraic Group Model (AGM)~\cite{fuchsbauer2018algebraic}. The AGM is used to analyze protocols that rely on discrete logarithm assumptions in prime-order groups, a fundamental aspect of Halo2's design. The AGM evaluates the security of cryptographic protocols by requiring that adversaries must explicitly compute group elements from previously observed elements, emphasizing discrete logarithm-type assumptions in prime-order groups.
\sysname~ guarantees completeness, soundness, knowledge soundness, and zero-knowledge properties under the AGM, assuming that all parties, including potential adversaries, are computationally bounded to probabilistic polynomial-time (PPT) algorithms. \sysname~ ensures that for any PPT adversary, there exists a PPT simulator such that, for any environment with arbitrary auxiliary input, the output distribution of the environment in a real-world execution (where a prover interacts with a verifier) is computationally indistinguishable from the output distribution in an ideal-world execution (where a simulator interacts with the verifier).

{\bf Oblivious circuits.} The proving logic of the circuits in \sysname~ is designed to be oblivious. This means that the execution of the proving algorithm does not depend on the specific values of the private inputs (the witness). Instead, the prover performs the same operations regardless of the actual witness values, which helps ensure that no information about the private inputs is leaked through the proof generation process or the resulting proof itself. For example, when implementing a sorting algorithm, the circuit would compare and swap elements in a fixed pattern regardless of their actual values, ensuring the same operations are performed for any input. Similarly, for conditional statements, both branches of the condition are typically evaluated, and their results are combined using a selector value, rather than following only one branch based on the private condition. To protect the privacy of table cardinalities and intermediate result sizes in join operations, we adopt the method introduced in ZKSQL~\cite{li2023zksql}. This approach involves introducing dummy tuples into our query evaluation process, effectively obscuring true data sizes while maintaining consistent row counts throughout query execution.

\subsection{Data Privacy Issues and Limitations}

\sysname~ sends the query results to the clients. Since the proof inherently includes the result of the query, a client could, for example, issue a query like "SELECT * FROM T" to attempt to retrieve all values from T, which threatens data privacy. Beyond the direct data exposure of returned results, sensitive data can also be exposed indirectly. Specifically, when data points or records in the database are correlated, query results may still leak sensitive information about records that are not part of the response.
For instance, correlation among records---such as similar attributes shared across groups or statistical dependencies---can allow a client to infer additional, sensitive details beyond the returned query results. In this way, even without direct access to records that are not included in query results, clients can potentially exploit patterns in the query response to gain insights into the broader dataset indirectly. This poses a data privacy risk that needs to be handled carefully.

\begin{sloppypar}
To address this issue, differential privacy techniques ~\cite{dwork2006differential, johnson2018towards}~ could be employed to ensure that individual data points are not revealed
either directly or indirectly. 
In this work, we do not incorporate differential privacy and leave it as an avenue of future work. We note, however, that methods of incorporating differential privacy to ZKP systems such as {\sysname} would lead to additional overhead in the circuit design.
\end{sloppypar}

\section{Custom Gates}
\label{sec:gates}
In this section, we present customized gates to represent SQL queries arithmetization in circuits.
%
%
Our goal is to introduce efficient designs by creating gates that have low-degree polynomials and a smaller number of circuit constraints. 
We design with low-degree polynomials because ZKP relies on cryptographic primitives where evaluating higher-degree polynomials is computationally expensive.

\vspace{-0.1in}
\subsection{Range Check}
\label{subsec:rangecheck}
We first introduce a range check gate as it is involved in many SQL operations like {\bf ``filter''}, {\bf ``sort''}, {\bf ``group by''} and {\bf ``join''}. 
%

Consider the range check statement $x \leq t$, where $x$ is a private value and $t$ is the public query input. A naive encoding compares $x$ against each possible value using the polynomial equation: 
$\prod_{i=0}^{t}(x-i) = 0$.
The degree of this polynomial grows linearly with $t$, making proof generation and verification computationally infeasible for large $t$.
%
%
In this work, we leverage a \emph{lookup table}~\cite{gabizon2020plookup} circuit structure to design the range check gate. The intuition behind using lookup tables lies in the idea of precomputing and storing results for a range of possible inputs. 
Our work builds upon the widely used Plookup framework~\cite{gabizon2020plookup}, which is well-established in the ZKP domain. While there are alternatives, such as Jolt~\cite{setty2024unlocking}, we leave exploration of these methods for future work.
Instead of reinventing cryptographic primitives, we focus on solving concrete implementation challenges to enhance efficiency of SQL operations within the PLONK framework. While previous works~\cite{zhang2017vsql, li2023zksql} implement operations like filter, sort, and join using boolean or logic gate frameworks, our approach utilizes arithmetic-based PLONKish circuits. This fundamental difference introduces significant challenges, as arithmetic circuits require different optimization strategies compared to boolean circuits. Adapting SQL operators in this context necessitates a  rethinking of their expression and optimization to achieve maximum efficiency.


\textbf{Definition.} The input to a range check gate, which checks if elements are between \( y_1 \) and \( y_m \), consists of a column \( C_{\text{in}} \) containing \( n \) elements \( \{x_1, x_2, \ldots, x_n\} \) to be range-checked, and a column (lookup table) \( T \) containing \( m \) sorted values \( \{y_1, y_2, \ldots, y_m\} \) representing the valid range, where \( y_1 < y_2 < \cdots < y_m \). The output of the range check gate is a column \( C_{\text{out}} \) containing \( n \) elements \( \{z_1, z_2, \ldots, z_n\} \), where each \( z_i \) indicates whether \( x_i \) is within the range defined by the lookup table \( T \). Specifically, \( z_i = 1 \) if \( x_i \in \{y_1, y_2, \ldots, y_m\} \), and \( z_i = 0 \) otherwise.

\textbf{Design A: A single range check.} We start with a simple case where we only want to prove that a single value $x$ is in a specific range $[0,t]$, hence proving that $0 \leq x \leq t$. 
%
In the first step of constructing the circuit, we create a private array $P$ (stored in an advice column) with the same length of set $Q$ where the first element in $P$ is $x$ and the other elements are any values copied from $Q$ (these values can be duplicates). Then, the prover supplies a permutation of $P$, denoted by $P'$ where $P'$ is private and stored in an advice column in the circuit. The values in $P'$ are sorted so that duplicate values are row-adjacent to each other.


In the second step, we establish a fixed column to store the set 
$Q$, arranging the values in ascending order. This organization ensures that both the prover and the verifier are aware of the values and their corresponding indices within the table. 
Such knowledge is crucial for determining the size of the lookup table needed.

%
The prover supplies a permutation of $Q$, denoted by $Q'$, where $Q'$ is private and stored in an advice column in the PLONKish circuit. The purpose of the permutation $Q'$ is to hide the position of values in $Q$ as the verifier knows all the information about $Q$. Then, the circuit can compare values in $Q'$ and $P'$ to check whether $x$ (in $P'$) is equal to some value in $Q'$, without revealing which value is that in $Q'$. 

Now, we show how we can check whether the value $x$ in $P'$ is equal to a value in $Q'$. The values in $Q'$ are arranged in a specific order such that either $P'_i = Q'_i$ or that $P'_i = P'_{i-1}$ where $P'_i$ and $Q'_i$ represent the $i$-th elements of $P'$ and $Q'$ respectively. 
Specifically, we enforce that the first values of $P$ and $Q$ are equal, i.e, $P'_i = Q'_i$ with $i=0$. If $P'_i \neq Q'_i$ for $i>0$, we enforce that $P'_i = P'_{i-1}$ meaning that $P'_i$ must be a duplicate of $P'_{i-1}$ in this case. Therefore, these constraints guarantee that 
every value in $P'_i$ is equal to some value in $Q'$. 
Formally, we enforce that either $P'_i = Q'_i$ or that $P'_i = P'_{i-1}$, using the rule:

\vspace{-0.1in}
\begin{equation}
\label{eq:c1}
    0 =
    \begin{cases}
        (P'_i - Q'_i) \cdot (P'_i - P'_{i-1}) &  \text{if } 1 \leq i \leq len(Q')-1, \\
        ~P'_i - Q'_i  & \text{if } i = 0.
    \end{cases}
\end{equation}
With the above polynomial constraints, the verifier knows that all the values of $P'$ are in $Q'$ without knowing the position information of the values in $P'$ and $Q'$ (therefore it does not know the exact values at each position in $P'$, preserving the privacy of the $x$ value).

Since $P'$ and $Q'$ are permutations of $P$ and $Q$, the polynomial constraints above ensure that all the values of $P$ (and $P'$) are in $Q$. To ensure this property, we develop polynomial constraints to ensure that both $P'$ is a permutation of $P$ and $Q'$ is a permutation of $Q$:
\begin{equation}
\label{eq:permutation}
     \prod_{i=0}^{len(Q)-1} (P_i + \alpha) (Q_i + \beta) = \prod_{i=0}^{len(Q)-1} (P'_i + \alpha) (Q'_i + \beta)
\end{equation}
where $P_i$, $Q_i$, $P'_i$ and $Q'_i$ represent the $i-$th element in $P$, $Q$, $P'$, and $Q'$ respectively, and $\alpha$ and $\beta$ are randomly chosen parameters.
The random values 
\(\alpha\) and \(\beta\) conceal the contents of the columns, ensuring confidentiality during verification and preventing zero products from causing collisions due to poor randomness.
%
%
%
To make the circuit formulation efficient, we express it as a recursive function to ensure that each equation maintains a low polynomial degree: 
\begin{equation}
\begin{gathered}
\label{eq:permu1}
    Z_{i+1} = Z_{i} \cdot \frac{(P_i + \alpha)(Q_i + \beta)}{(P'_i+\alpha)(Q'_i + \beta)}\\
    Z_{\text{len}(Q)} = Z_0 = 1
\end{gathered}
\end{equation}

\begin{figure}[t]
\centerline{\psfig{figure=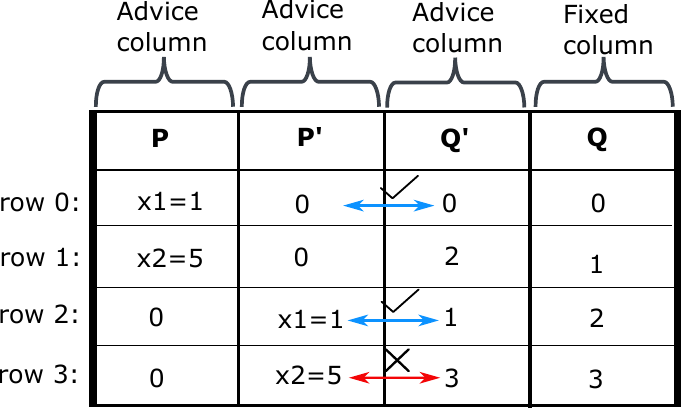, width=0.3\textwidth} }
\vspace{-0.1in}
\caption{Range check with lookup tables illustration. }
\label{fig:lookup}
\vspace{-0.2in}
\end{figure}

\begin{example}
\label{example:1}
Figure~\ref{fig:lookup} illustrates our proposed lookup table circuit design for proving a range check statement without revealing the specific values of $x_1$ and $x_2$. 
The prover uses a fixed column \( Q \) accessible to verifiers, containing values in the range \([0, 4)\). An advice column \( P \) includes the actual values of \( x_1 \) and \( x_2 \) with the remaining cells filled arbitrarily from \( Q \). The prover then generates advice columns \( P' \) and \( Q' \) as permutations of \( P \) and \( Q \), respectively, with \( P' \) sorted in ascending order and duplicates adjacent, ensuring \( Q' \) starts with the same value as \( P' \).
The prover checks that each \( P'_i \) is either equal to \( Q'_i \) or \( P'_{i-1} \). This polynomial constraint ensures that values are within the range. 

\end{example}

\textbf{Design B: Batching range check.} The lookup table technique facilitates verifying whether all elements of an array \(P\) belong to the set \(Q = \{i \mid 0 \leq i \leq t, i\in \mathbb{Z}\}\), effectively checking if they fall within the range \([0, t]\). By organizing the elements of \(P\) into a single advice column and aligning the set \(Q\) within a fixed column, the lookup table approach previously discussed can be applied. 
For instance, the circuit depicted in Figure~\ref{fig:lookup} can demonstrate whether the elements in the array \([0,2,1,3]\) (column \(Q'\)) are contained within the range \([0, 4)\) (column \(Q\)).
The complexity of verifying the inclusion of the arrays \([0]\) and \([0,1,2,3]\) within the range \([0, 4)\) remains consistent, as it necessitates the application of formulas~\ref{eq:c1} and \ref{eq:permu1} for each row. 
%

{\bf Design C: Optimizing range checks with bitwise decomposition and lookup tables.}
To mitigate the scalability concerns associated with range checks, especially when the size of the set \(Q\) becomes significantly large (e.g., \(2^{64}\)), we propose an optimization technique that leverages bitwise decomposition in conjunction with lookup tables. This method entails representing integers in a fixed bit-length format, such as 64 bits, and subsequently verifying the integrity of this bitwise representation in relation to the original integer value.

Given an integer \(N\) and its representation as a sequence of bits \(b_0, b_1, \ldots, b_{k-1}\), where \(b_0\) is the least significant bit and \(b_{k-1}\) the most significant, the relationship between \(N\) and its bits is described by $N = \sum_{i=0}^{k-1} b_i \cdot 2^i$.
%
%
We partition the bitwise representation of integer \(N\) into smaller segments of 8 bits, referred to as u8 cell. This approach is predicated on the standard binary representation of integers, wherein the integer is segmented into 8-bit blocks. For instance, a 64-bit integer is divided into 8 u8 cells, each encapsulating an 8-bit slice of the integer. Constraints are imposed to ensure this decomposition is accurately executed with the constraints $N = \sum_{i=0}^{7} c_i \cdot 2^{8i}$ for a 64-bit integer $N$,
where $c_i$ represents the 8-bit segment of $N$ at position $i$.
Each u8 cell $c_i$ should have values within the range $0$ to $255$ (inclusive), corresponding to $2^8 - 1$. To validate each u8 cell, we utilize a fixed-sized lookup table of size 256, which contains integers from $0$ to $255$. This table allows efficient range checking for each segment of the integer.
The advantage of using this lookup table is that it is fixed-sized (256 entries) and can be reused multiple times for each u8 cell check. This ensures that each u8 cell $c_i$ can quickly verify if its value falls within the allowed range of \(0\) to \(2^8 - 1\).
By reusing the lookup table across all 8 u8 cells of $N$, we streamline the validation process and ensure consistent range checks.



{\bf Design D: Conditional statements proving.}
The previous range check method faces limitations when a value falls outside the lookup range, making proof construction difficult. 
%
To this end, we introduce an augmented method that seamlessly integrates with lookup tables to facilitate range checks while gracefully handling conditional scenarios. 
%
When dealing with data filtering, an upper-bound value, denoted as $u$, typically exists. Consequently, for \(x \geq t\), this also implies \(u > x \geq t\). Therefore, proving \(u > x \geq t\) is equivalent to proving 
$0 \leq x - t \leq u$.
%
%
To establish \(x < t\), it suffices to demonstrate that \(x - t < 0\). By adding \(u\) to both sides of the inequality, we obtain \(0 \leq x - t + u < u\). Introducing a binary variable, denoted as $check$, to determine whether \(x < t\), we ultimately prove the following statement:
\begin{equation}
\label{eq:lessthan2}
    0 \leq (x - t) + \text{{check}} \cdot u < \text{{u}}
\end{equation}
To accomplish this, the prover configures several supporting columns in the PLONKish circuits. Initially, an advice column is created to output \(1\) if \(x < t\) and \(0\) otherwise. Additionally, another advice column is established to store the values of \(x - t\). Finally, the prover undertakes the task of proving that \(\text{{check}} + (x - t)\) falls within the range of \([0, u)\) with the assistance of the lookup tables introduced in Section~\ref{subsec:rangecheck}. 


Note that the values in the \textbf{``check''} columns are prover-determined, with no explicit constraints imposed among \(x\), \(t\), and \textbf{``check''}. However, if the \textbf{``check''} values are inaccurately provided, the proof generation process encounters a failure.
Moreover, the values in the \textbf{``x''}, \textbf{``t''}, and \textbf{``x-t''} columns adhere to the constraints \(cell_i(x) - cell_i(t) - cell_i(x-t) = 0\) for the initial four rows, where \(cell_i(x)\) represents the value in the \textbf{``x''} column at row \(i\). These constraints guarantee that the discrepancies between corresponding elements in the \textbf{``x''} and \textbf{``t''} columns align with the values stored in the \textbf{``x-t''} column.




\textbf{Correctness.}
We prove the correctness of the range check gate from two aspects.
(1) Property 1: Element Inclusion. All the values in \( P' \) are in \( Q' \). Assume that there exists one value \( x \) in \( P' \) that is not in \( Q' \). We will show that this assumption leads to a contradiction. If \( x \) is the first value in \( P' \), by Equation~\eqref{eq:c1}, \( 0 = P'_0 - Q'_0 \), implying \( P'_0 = Q'_0 \). Thus, \( x = P'_0 \) contradicts \( x \notin Q' \). If \( x \) is not the first, for \( 1 \leq i \leq \text{len}(Q') - 1 \), \( 0 = (P'_i - Q'_i)(P'_i - P'_{i-1}) \) enforces \( P'_i = Q'_i \) or \( P'_i = P'_{i-1} \). If \( P'_i = Q'_i \), \( x \in Q' \); if \( P'_i = P'_{i-1} \), \( x \) duplicates \( P'_{i-1} \), tracing back to \( P'_0 = Q'_0 \). Thus, \( x \) must be in \( Q' \), contradicting \( x \notin Q' \). Hence, all \( P' \) values are in \( Q' \), proving Equation~\eqref{eq:c1}'s constraints ensure \( P' \subseteq Q' \).
(2) Property 2: Permutation Integrity. Assume \( P' \) and \( Q' \) are not permutations of \( P \) and \( Q \). This implies that some \( P_i \) or \( Q_i \) does not match any corresponding \( P'_i \) or \( Q'_i \). Since the products involve symmetric polynomials, if \( \{P'_i + \alpha\} \cup \{Q'_i + \beta\} \) are not permutations of \( \{P_i + \alpha\} \cup \{Q_i + \beta\} \), the two sides of the equation cannot be equal due to the unique factorization of polynomials. Therefore, by the Fundamental Theorem of Symmetric Polynomials, \( P' \) and \( Q' \) must be permutations of \( P \) and \( Q \).


With the two properties proven above, we can guarantee the correctness of proving \( x < t \) if \( x \) is not larger than \( t \).
The inequality in Equation~\ref{eq:lessthan2} holds if \( check = 1 \) and \( x < t \), or if \( check = 0 \) and \( x \leq t \). Therefore, setting the binary variable \( \text{check} \) correctly is sufficient to determine whether \( x < t \) or not. By transforming \( x \) to \( (x - t) + \text{check} \cdot u \), we can guarantee that the transformed \( x \) is in the range of \( 0 \) to \( u \). 


\textbf{Complexity of a Range Check Gate.} The ZK proof generation cost depends on the number of constraints. We analyze the complexity of a gate by specifying the required number of each type of constraint.
To validate that all elements of an array \(P\) reside within a set \(Q\), the number of constraints, as defined in Equations~\ref{eq:c1} and~\ref{eq:permu1}, corresponds to the greater of \(|P|\) or \(|Q|\), denoted as \(\max(|P|, |Q|)\).
When employing the bitwise decomposition method for an input set \(P\) where \(|P| > 256\), the lookup table is padded to match the size of the input by including duplicates of values ranging from 0 to 255.
For 64-bit integers decomposed into 8 u8 cells, each cell undergoes a range check utilizing the lookup table. The number of constraints, as described by Equations~\ref{eq:c1} and~\ref{eq:permu1}, is equivalent to \(8 |P|\).
Additionally, the number of constraints required to ensure the correct decomposition of integers into u8 cells, as well as verifying that all segments of the integer fall within the specified ranges, is \(|P|\).
Finally, to transform a value \(x\) with \(x' = (x - t) + \text{check} \cdot u\) (see Equation~\ref{eq:lessthan2}), an additional \(|P|\) constraints are needed.

\vspace{-0.1in}
\subsection{Sort}
\label{subsec:sort}
We detail our approach to proving the correctness of sort operations. 



\textbf{Definition.} The input of a sort gate consists of a table \( D \) with \( m \) columns \( C_1, C_2, \ldots, C_m \), where \( C_i \) represents different attributes or values to be sorted. Each column \( C_i \) contains \( n \) elements \( \{x_{i1}, x_{i2}, \ldots, x_{in}\} \). 
The output is a table \( SD \) with \( m \) columns \( C'_1, C'_2, \ldots, C'_m \), where each column \( C'_i \) contains the sorted elements \( \{y_{i1}, y_{i2}, \ldots, y_{in}\} \). Here, \( y_{ij} \) represents the \( j \)-th element in column \( C'_i \) of \( D \), arranged in the desired order (e.g. increasing or decreasing).


\textbf{Design.} The first step in proving sort operations entails generating a witness, which includes the result of applying a specific sorting algorithm to the input data. The prover has the flexibility to choose any sorting algorithm, as long as the resulting order is correct.
%
Let $D$ and $R$ represent the input data and the sorted result, respectively. Two essential properties of $R$ must be guaranteed for a successful proof. First, the data in $R$ should match that in $D$ except for the order, leading to a permutation check between $R$ and $D$. The following constraints, akin to Equations~\eqref{eq:permu1}, ensure this:
\vspace{-0.1in}
\begin{equation}
\begin{gathered}
\label{eq:permu2}
    Z_{i+1} = Z_{i} \cdot \frac{R_i + \alpha}{D_i+\alpha}\\
    Z_{\text{len}(D)} = Z_0 = 1
\end{gathered}
\end{equation}
Here, $D_i$ and $R_i$ represent the $i$-th element in $D$ and $R$, and $\alpha$ is a randomly chosen parameter similar to ones in Equation~\ref{eq:permutation}.

In sorting mechanisms where multiple attributes are considered, a unified approach is adopted to encapsulate these attributes into a singular composite entity. This is achieved by allocating a consistent bit-length representation for each attribute, specifically employing a 64-bit format for this purpose. Such a fixed bit-length representation is critical in preserving the intrinsic value hierarchy and relative ordering of each attribute during the process of concatenation. 
%
%
In addition, the data in $R$ must align with the sort definition. To verify this, we check that $R_i \leq R_{i+1}$ (assuming an ascending order) for $i$ in the range $[0, ~\text{len}(R)-1)$. This is achieved by proving the transformed statement introduced in Equation~\ref{eq:lessthan2} with the assistance of lookup tables. 
%

\textbf{Correctness.} We prove the correctness of the two properties introduced in the sort gate.
(1) Property 1: Permutation Integrity. \( R \) is a permutation of \( D \). The proof follows a similar approach used to prove that \( P' \) is a permutation of \( P \). For details, refer to the proof in Section~\ref{subsec:rangecheck}. 
(2) Property 2: Sortedness. \( R \) is sorted in ascending order. Assume, for contradiction, that \( R \) is not sorted in ascending order. This implies there exists at least one pair of indices \( i < j \) such that \( R_i > R_j \).
Since we enforce the constraints \( R_i \leq R_{i+1} \) for each pair \( (R_i, R_{i+1}) \) where \( i \in [0, \text{len}(R)-1) \), and the correctness of the range check gate for \( R_i \leq R_{i+1} \) is proven in Section~\ref{subsec:rangecheck}, we guarantee that \( R_i \leq R_{i+1} \) holds for all valid indices \( i \).
This contradicts the assumption that there exists at least one pair of indices \( i < j \) such that \( R_i > R_j \). Therefore, \( R \) must indeed be sorted in ascending order to satisfy the defined properties and constraints of the sort operation.


\textbf{Complexity of a sort gate.}
Permutation checks between $D$ and $R$ require \(|D|\) constraints, as per Equation~\ref{eq:permu2}. Additionally, checking $R_i \leq R_{i+1}$ for $i \in [0, \text{len}(R)-1)$ requires constraints proportional to \(|D|\), similar to the range check gate in Section~\ref{subsec:rangecheck}.

\vspace{-0.1in}
\subsection{Group-by}
\label{sec:groupby}
In this section, we outline our methodology for verifying the correctness of group-by operations.

\textbf{Definition.} The input of a group-by gate consists of a table \( D \) with \( n \) columns \( C_1, C_2, \ldots, C_n \), where each column represents different attributes of \( D \). 
The output is a table \( SD \) with \( n + 2 \) columns that rearranges the records of \( D \) such that records with identical values on the grouping attributes \( \mathit{G} \) are placed into the same group-by bin. Alongside the original \( n \) columns from \( D \), \( SD \) includes two additional columns:
(1) \( \text{start\_index} \): Indicates the starting index of each group-by bin in \( D \).
(2) \( \text{end\_index} \): Indicates the ending index of each group-by bin in \( D \).
The \( \text{start\_index} \) and \( \text{end\_index} \) are used for the later aggregation functions such as SUM and others.

{\bf Design.} 
%
Given an input table $D$ and the group-by attributes $\mathit{G}$, the prover first generates the sorted table $SD$ based on the group-by attributes $\mathit{G}$. This sorting ensures that records with identical values in $\mathit{G}$ are adjacent in $SD$. To verify that $SD$ is a sorted version of $D$, we employ the approach introduced in Section~\ref{subsec:sort}.

To identify the starting and ending indices of each group-by bin, we check each record in $SD$ on the group-by attributes $\mathit{G}$. We use a binary value $b$ to indicate whether a record in $SD$ is a starting or ending record, where $1$ signifies that the record is either a starting or ending record. The constraint to check whether two values $v_1$ and $v_2$ are equal or not with a binary value $b$ is as follows:
\begin{equation}
\label{eq:iszero}
    b = 1 - (v_1 - v_2) \cdot p
\end{equation}
where $p$ is the value provided by the prover. Specifically, $p = 0$ if $v_1 = v_2$ and $p = \frac{1}{v_1 - v_2}$ otherwise. To ensure that the prover provides the correct value of $p$, we add the following constraint for each pair of $v_1$ and $v_2$:
\begin{equation}
\label{eq:constraint}
    b \cdot (v_1 - v_2) = 0
\end{equation}
%
A record is marked as the start of its bin if no preceding adjacent records share identical values in group-by attributes $\mathit{G}$. Conversely, it is marked as the end of its bin if no subsequent adjacent records share identical values in $\mathit{G}$.




\textbf{Correctness.} 
We utilize the sort gate introduced in Section~\ref{subsec:sort} to ensure that \( SD \) is sorted. The correctness of the sort gate is demonstrated in Section~\ref{subsec:sort}.
Next, we deduce the correctness of the starting and ending indices of each group-by bin.
%
%
According to Equation~\ref{eq:iszero}, if $v_1 = v_2$, then $b = 1$, and Equation~\ref{eq:constraint} holds trivially as $b \cdot (v_1 - v_2) = 0$. If $v_1 \neq v_2$: (1) If $p = \frac{1}{v_1 - v_2}$, then $b = 0$, and Equation~\ref{eq:constraint} holds because $b \cdot (v_1 - v_2) = 0$. (2) If $p = 0$, then $b = 1$. In this case, Equation~\ref{eq:constraint} does not hold as $b \cdot (v_1 - v_2) = (v_1 - v_2) \neq 0$. Consequently, a valid proof cannot be generated because the system detects the inconsistency. 


%
%

\textbf{Complexity of a group-by gate.} 
The group-by operation is facilitated through the sorting of associated attributes, followed by a verification of the accuracy of sorting. 
Consequently, the computational complexity and the number of constraints required for ensuring the correctness of group-by operations are identical to those identified in the analysis of sorting constraints. Additionally, the number of constraints (Equations~\ref{eq:iszero} and ~\ref{eq:constraint}) needed to identify the starting and ending indices of each group-by bin is $2$\(|D|\) where $D$ is the input set.

\begin{figure}[t]
\centerline{\psfig{figure=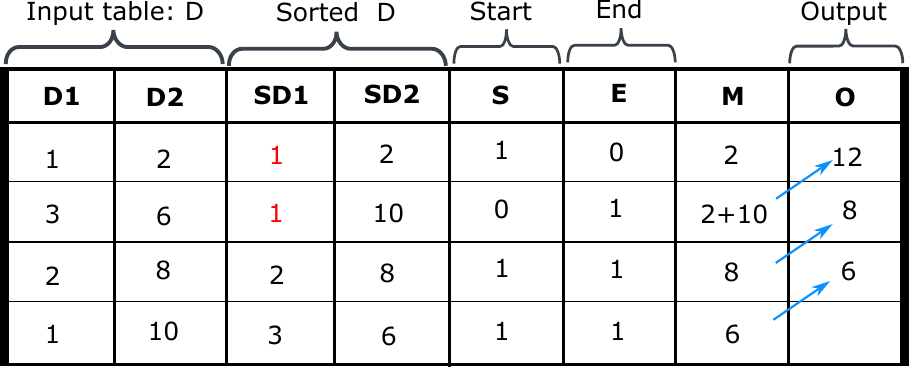, width=0.35\textwidth} }
\vspace{-0.1in}
\caption{Illustration with the SQL query ``SELECT SUM(D2) FROM {\bf T} GROUP BY D1.''. }
\label{fig:groupby}
\vspace{-0.2in}
\end{figure}

\begin{example}
    \label{example:groupby}
    Figure~\ref{fig:groupby} illustrates the SQL query "SELECT SUM(D2) FROM {\bf T} GROUP BY D1." The input table has columns ${\bf D1}$ and ${\bf D2}$. The prover adds sorted columns ${\bf SD1}$ and ${\bf SD2}$, as well as columns ${\bf S}$ and ${\bf E}$ for group-by bin indices.
    An advice column ${\bf M}$ is created where $M_i$ is defined as $M_{i-1} + SD2_i$ if $SD1_i = SD1_{i-1}$, otherwise $M_i = SD2_i$, with $M_0 = SD2_0$.
    The final output column ${\bf O}$ captures the result by copying only the last record of each group-by bin, as indicated by the ${\bf E}$ column.
\end{example}




\vspace{-0.1in}
\subsection{Join}
\label{section:join}
Now, we present our methodology for verifying the correctness of join operations.

\textbf{Definition.} The input of a join gate consists of two tables \( T1 \) and \( T2 \), each represented by \( m \) columns \( C_{1}^{(1)}, C_{2}^{(1)}, \ldots, C_{m}^{(1)} \) for \( T1 \) and \( n \) columns \( C_{1}^{(2)}, C_{2}^{(2)}, \ldots, C_{n}^{(2)} \) for \( T2 \). These columns correspond to different attributes of \( T1 \) and \( T2 \), respectively. The joining operation is performed based on equality conditions between specified joining attributes \( \mathit{J1} \) from \( T1 \) and \( \mathit{J2} \) from \( T2 \), which may involve any of the columns \( C_i^{(1)} \) from \( T1 \) and \( C_j^{(2)} \) from \( T2 \). The output is a table \( JD \) that combines records from \( T1 \) and \( T2 \) where the joining condition \( T1.\mathit{J1} = T2.\mathit{J2} \) is satisfied. 
The table \( JD \) has \( m+n \) columns:
(1) The first \( m \) columns \( C_{1}^{(1)}, C_{2}^{(1)}, \ldots, C_{m}^{(1)} \) are the attributes from \( T1 \).
(2) The next \( n \) columns \( C_{1}^{(2)}, C_{2}^{(2)}, \ldots, C_{n}^{(2)} \) are the attributes from \( T2 \).
%

\textbf{Design.}
Consider two private tables $T1$ and $T2$, and let $p$ be the join predicate. 
%
To establish the correctness of the join operations, the prover calculates several types of witnesses locally. Initially, the prover creates two new tables $T'1$ and $T'2$ to reorder the records within them. Specifically, each table $T1'$ and $T2'$ is split into two parts: $T1'_{p}$ and $T2'_{p}$ contain records contributing to the join predicate, while $T1'_{non-p}$ and $T2'_{non-p}$ contain records that do not contribute to the join predicate. The correctness of this reordering is verified using polynomial constraints, as detailed in Equation~\ref{eq:permu2}.
%

Next, the prover ensures that the records in $T1'$ and $T2'$ indeed contribute to the join predicate. 
In the context of primary key-foreign key joins (non-primary key-foreign key joins are discussed later), assuming that table T1 contains the foreign key and table T2 contains the primary key,
instead of checking the existence of records in one table in another, the prover proves that $T2'_{non-p}$ is disjoint from $T1'_{non-p}$ and $T1'_{p}$. 
%
Traditional straight-forward methods involve checking each value in $T1'_{non-p}$ against $T2'_{non-p}$, resulting in a large number of polynomial constraints. To address this, a sorted table $S$ is created to store the unique values in $T1'_{non-p}$ and $T2'_{non-p}$, with a proof of $S_i < S_{i+1}$ for $i \in [0, \text{len}(S)-1)$. 

However, duplicates~\footnote{Since SQL operates on multisets, duplicates are not only a concern in join algorithms but also throughout the entire query process, as the result must accurately reflect the correct number of duplicates.
In Plookup~\cite{gabizon2020plookup}, this can be efficiently managed by encoding the frequency of elements in the multiset into polynomial commitments. Plookup then ensures that the correct number of occurrences for each value is preserved during query execution, maintaining the semantic integrity of SQL operations.} in $S$ pose challenges in determining their origin (from the same or different tables). To resolve this, the prover implements a deduplication strategy through the creation of distinct versions, namely $T1^{de}$ and $T2^{de}$. This process ensures that each value in $T1'{non-p}$ is accounted for in $T1^{de}$ and every value in $T2'{non-p}$ is accounted for in $T2^{de}$.
To establish this deduplication property, we leverage the mechanism of lookup tables we proposed in Section~\ref{subsec:rangecheck}. The verification of the range check operation, which confirms the existence of each value in a column within a lookup table (i.e. another column), is adapted to ensure the deduplication of $T1^{de}$ and $T2^{de}$. It is noteworthy that the distinction between $T1^{de}$, $T2^{de}$, and a lookup table for range check lies in their roles within the PLONKish circuit configuration. Specifically, $T1^{de}$ and $T2^{de}$ serve as \texttt{advice columns}, signifying the privacy of the data they contain, while a lookup table for range check is stored in an \texttt{instance column}, designating the public nature of the data within it. This distinction is integral to the overall architecture of the PLONKish circuit configuration.
The prover establishes that $S$ is a permutation of $T1^{de} \cup T2^{de}$, ensuring that $S$ matches the records in $T1^{de}$ and $T2^{de}$ except for the order.  And $S_i<S_{i+1}$ holds for $\forall i \in [0, len(S)-1)$.

\begin{figure}[t]
\centerline{\psfig{figure=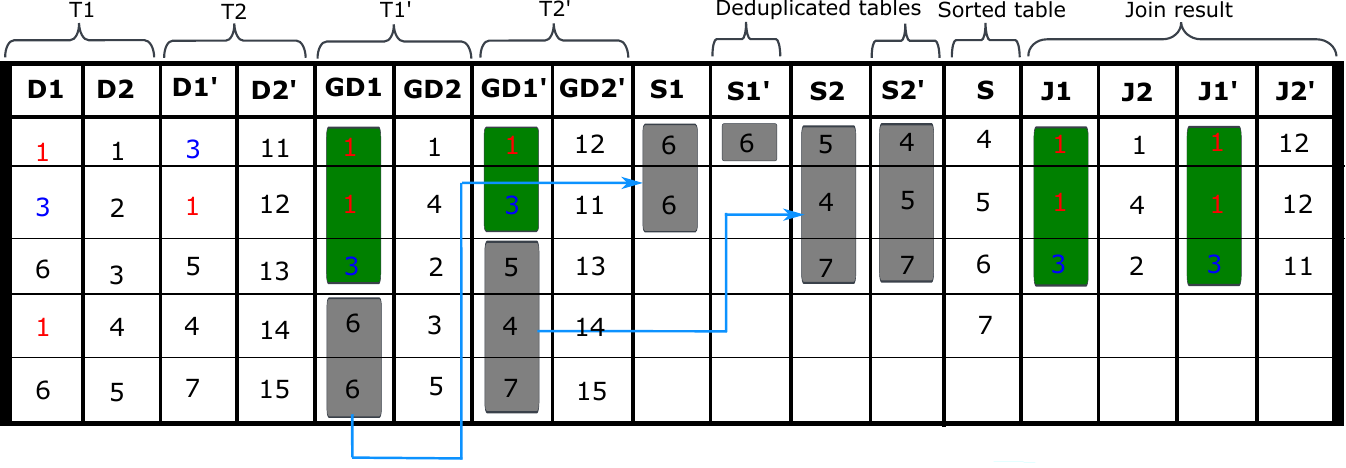, width=0.5\textwidth} }
\vspace{-0.1in}
\caption{Illustration with the SQL query ``SELECT T1.D1, T2.D2' FROM {\bf T1, T2} WHERE  T1.D1 = T2.D1' ''. }
\label{fig:join}
\vspace{-0.3in}
\end{figure}

Next, we introduce the method for generating join results from contributing records.
%
Consider the scenario where the join predicate between two tables, \(T1\) and \(T2\), is defined as \(T1.\text{attr1} = T2.\text{attr2}\). In the context of primary key-foreign key joins, the uniqueness of primary keys implies the absence of duplicates, ensuring that each foreign key in \(T1\) corresponds to at most one matching row in \(T2\). Let us denote \(T1'_p\) and \(T2'_p\) as the subsets of \(T1\) and \(T2\), respectively, that are relevant to the join predicate. Within \(T1\), \(\text{attr1}\) serves as the foreign key, whereas \(\text{attr2}\) is the primary key within \(T2\).
For each record in \(T1'_p\), a search is conducted within \(T2'_p\) to identify any record that satisfies the join predicate. Upon finding a match, the record from \(T2'_p\) is concatenated with the corresponding record from \(T1'_p\), forming a combined record.
To verify the join results, two key properties must be established:
(1) \textbf{Equality Verification}: For each concatenated record \(r\), we set the polynomial \(r.\text{attr1} - r.\text{attr2} = 0\), where \(\text{attr1}\) and \(\text{attr2}\) are the attributes from \( T1 \) and \( T2 \), respectively, used for the join. 
(2) \textbf{Source Verification}: To ensure all records joined with \(T1'_p\) originate exclusively from \(T2'_p\), we use lookup tables as described in Section~\ref{subsec:rangecheck}.


To handle joins without relying on primary key-foreign key relationships, we compute the join result by pairwise comparing records from \( T1'_p \) and \( T2'_p \) using a similar procedure. 
However, the method introduced above may not include all records contributing to the join predicate in \( T1' \) and \( T2' \). To ensure completeness, we additionally prove that \( T1'_{non-p} \) is disjoint from \( T2' \) using the same deduplication and sorting mechanism.
This ensures that all relevant records are considered in the join operation.

\textbf{Scalability.}
%
When managing a large number of join operations, the database's query engine
typically executes these joins sequentially, joining
two tables at a time. This results in various execution plans having
differing numbers of total and intermediate join results.
Optimizations aimed at reducing the number of these total and intermediate join results should be identified and applied prior to the circuit design phase. 

\textbf{Correctness.} 
To ensure the correctness of the join gate, we verify the following properties:
(1) Property 1: Permutation Integrity. The sets \( T1'_{p} \cup T1'_{non-p} \) and \( T2'_{p} \cup T2'_{non-p} \) are permutations of the original tables \( T1 \) and \( T2 \), respectively. The proof follows a similar approach used to prove that \( P' \) is a permutation of \( P \). For details, refer to the proof in Section~\ref{subsec:rangecheck}. 
(2) Property 2: Completeness. \( T1' \) and \( T2' \) include all records that contribute to the join predicate. Assume \( T1' \) misses a record \( x \) that contributes to the join predicate, i.e., \( x \) is in \( T1'_{non-p} \). We prove that \( T1'_{non-p} \) is disjoint from \( T2' \) and \( T2'_{non-p} \) using the Element Inclusion and Sortedness properties as detailed in Section~\ref{subsec:sort}. This implies \( x \) does not overlap with \( T2' \), indicating that \( x \) does not contribute to the join predicate. This contradicts the assumption, proving that \( T1' \) includes all relevant records. Similarly, \( T2' \) includes all records that contribute to the join predicate.
(3) Property 3: Exclusivity. \( T1' \) and \( T2' \) do not include records that do not contribute to the join predicate. Assume \( T1' \) includes a record \( y \) that does not satisfy the join condition. During the join process, the concatenated record \( r \) involving \( y \) would fail the constraint \( r.\text{attr1} - r.\text{attr2} = 0 \), contradicting the assumption. Therefore, \( T1' \) and \( T2' \) only include records that contribute to the join predicate.
The correctness of the above three properties ensures the correctness of the join gate.

\begin{example}
    Figure~\ref{fig:join} illustrates the SQL query "SELECT T1.D1, T2.D2 FROM T1, T2 WHERE T1.D1 = T2.D1'."
    Upon receiving the input tables \(T1\) and \(T2\), the prover creates new tables \( T1' \) and \( T2' \). The upper parts contain records contributing to the join \( T1.D1 = T2.D1' \). The green area in columns \( GD1 \) and \( GD1' \) shows corresponding values.
    The prover verifies that the non-contributing records (gray areas) in \( GD1 \) and \( GD1' \) do not intersect by constructing columns \( S1 \) and \( S2 \), removing duplicates, and ensuring that each value in \( S1 \) and \( S2 \) exists in their respective sorted columns \( S1' \) and \( S2' \). The process of proving that the non-contributing records (gray areas) in \( GD1' \) do not intersect with the contributing records (green areas) in \( GD1 \) is performed in a similar manner. This step is omitted in the example and assumes that table \( T1 \) contains the foreign key and table \( T2 \) contains the primary key.
    A column \( S \) is constructed by sorting \( S1' \) and \( S2' \). The prover checks \( S \) is a permutation of \( S1' \cup S2' \) and that \( S_i < S_{i+1} \) for all \( i \).
    The join results, as depicted in the last four columns, are derived from the green-highlighted values in \( GD1 \) and \( GD1' \).
\end{example}

\textbf{Complexity of a join gate.}
Given two tables $T1$ and $T2$ with the number of records denoted by $T1_{\text{num}}$ and $T2_{\text{num}}$ respectively, we categorize records that contribute to the join predicate as $T1^{\text{join}}$ and $T2^{\text{join}}$, and those that do not contribute as $T1^{\text{disjoin}}$ and $T2^{\text{disjoin}}$. The counts of these records are denoted as $T1_{\text{join\_num}}$, $T2_{\text{join\_num}}$, $T1_{\text{disjoin\_num}}$, and $T2_{\text{disjoin\_num}}$ respectively. We omit the copy constraints in this analysis as they are lightweight.
%
A permutation or range check gate, sized at $X$, implies that $X$ corresponding constraints are applied across two columns, each populated with $X$ values. The computation of constraints for a join operation encompasses five distinct categories:

\begin{itemize}[leftmargin=*]
    \item Two permutation check gates 
    with sizes $T1_{\text{num}}$ and $T2_{\text{num}}$.
    \item Two range check constraints with lookup tables (referred to by Equations~\ref{eq:c1} and~\ref{eq:permu1}) in sizes $T1_{\text{disjoin\_num}}$ + $T2_{\text{num}}$ and $T2_{\text{disjoin\_num}}$ + $T1_{\text{num}}$ for the deduplication process.
    \item One range check constraint with lookup tables (referred to by Equations~\ref{eq:c1} and~\ref{eq:permu1}) in sizes $T1_{\text{disjoin\_num}}$ + $T2_{\text{num}}$ and $T2_{\text{disjoin\_num}}$ + $T1_{\text{num}}$ (in the worst case) for sorting the deduplicated versions.
    \item The number of equality check constraints (in the form $x-y=0$) for checking if corresponding values satisfy the join predicate, with a maximum of either $T1_{\text{disjoin}}$ or $T2_{\text{disjoin}}$, for columns such as $J1$ and $J1'$ in the given figure.
    \item One range check constraint with lookup tables (referred to by Equations~\ref{eq:c1} and~\ref{eq:permu1}) in the size of $\max(T1_{\text{disjoin}}, T2_{\text{disjoin}})$ to ensure all records joined with $T_2^{\text{join}}$ originate exclusively from $T_2^{\text{join}}$.
\end{itemize}
\vspace{-0.1in}

\subsection{Aggregation and Other Operations}
Since aggregation operations are often applied together with the group-by operation, we describe how to implement them in conjunction with group-by. The SUM gate is implemented by establishing a column that holds intermediate, non-final values for each group-by bin, as described in Figure~\ref{fig:groupby} and Example~\ref{example:groupby}. To identify the starting and ending indices of each group-by bin, we can follow the method introduced in the Group-by section~\ref{sec:groupby}.
Once we determine the indices of these boundary records for each group-by bin, we can employ a similar approach to implement the COUNT gate. With the SUM and COUNT values determined for each bin, the AVERAGE gate can be naturally realized through a division gate that processes these values.

Furthermore, the MAX and MIN gates are facilitated by a sorting mechanism. By arranging the values in ascending order, the smallest and largest values, corresponding to the MIN and MAX gates respectively, can be directly identified as the first and last values in the sorted list. Along with these functionalities, we have implemented additional aggregate functions such as Standard Deviation, Variance, and Median. Additionally, we have developed capabilities for string matching and concatenation by validating the equality of sub-strings in two strings using lookup tables.

For projection operation, we use selectors to project the desired columns by setting them to 1 for inclusion and 0 for exclusion. Each selector controls a multiplication gate, multiplying the column by 1 or 0 based on whether it is part of the projection. To ensure that the projection results include all the desired columns without revealing their positional information, we employ a lookup table technique. This approach guarantees that the output preserves the required columns while maintaining the privacy of their original positions.
%

The set operations can be implemented using the methods described for the join gate. Set equality is handled by first sorting both tables and then comparing tuples at each index. Set disjointness is checked by sorting both tables and ensuring that any consecutive tuples $R_i$ and $R_{i+1}$ in the sorted list satisfy $R_i \leq R_{i+1}$. Set intersection is achieved as illustrated in Example 4.3 and Figure~\ref{fig:join}, where the join method is applied to extract common tuples between tables 
$R$ and $S$. For set union, tuples that are common to both $R$ and $S$ (found via set equality) are first removed from $R$, and the remaining tuples are then concatenated to $S$.

We have covered the most common operations used in SQL queries. Other variations of these operations can be constructed using the methods introduced in this work, as long as they can be represented within the circuit framework.

\vspace{-0.1in}

\subsection{Combining Gates}
PoneglyphDB processes full SQL queries by combining customized gates for different operators as follows:
\begin{itemize}[leftmargin=*]
    \item [1.] \textbf{Mapping Operations to Gates:} Each SQL query operation, such as sorting or joining, is represented by a corresponding gate. This gate executes the specific operation, ensuring accurate relationships between inputs and outputs.
    
    \item [2.] \textbf{Predefined Execution Plan and Assembly:} The SQL query's predefined execution plan outlines the sequence and dependencies of operations, guiding the assembly of gates in sequence. 
    Each gate's output serves as the input for the next, ensuring data flows correctly through the circuit according to the optimized plan.
    
    \item [3.] \textbf{Combining Gates:} Multiple gates are combined to handle various operations as outlined in the execution plan. The gates are strategically assembled in sequence, with the output of one gate serving as the input for the next.
    
\end{itemize}
Since each operator (such as sorting or aggregation) is verified separately, proving each ensures the correctness of the entire query. However, even when all inputs appear in the output, as with sorting, there is a risk of data leakage from intermediate steps like comparisons. This is because intermediate steps, like comparing the relative order of data elements, may reveal unintended patterns or relationships. This is why we implement oblivious circuits: to ensure that no information about intermediate steps, such as comparisons or data order, is exposed during proof generation.

\textbf{Correctness.} 
Let \( G_1, G_2, \ldots, G_n \) be a sequence of gates processing a query. Assume \( G_1 \) receives correctly transformed input and operates correctly, yielding correct output \( O_1 \). 
Assume for each \( G_i \) (where \( 1 \leq i \leq n \)), the output \( O_i \) is correct and serves as input to \( G_{i+1} \). Given each gate's correctness, \( G_{i+1} \) operates correctly on \( O_i \) to produce \( O_{i+1} \).
By induction, the final output \( O_n \) produced by the last gate \( G_n \) is correct.

\section{EXPERIMENTAL RESULTS}
\label{sec:evaluation}
We evaluate \sysname~
%
%
in terms of proving and verification time, memory usage, operator performance, proof size and scalability.

\vspace{-0.1in}
\subsection{Experimental Setup}

We implement \sysname's circuits and gates using Halo2, a state-of-the-art ZKP system~\cite{halo2}. The implementation of all Plonkish circuits for the queries is conducted using Rust. Our evaluation of \sysname~ focuses on a selected subset of the TPC-H benchmark~\cite{TPC-H2023}, specifically targeting queries that are representative of common data analytics workloads. 

We compare with ZKSQL~\cite{li2023zksql}, a state-of-the-art solution for using interactive ZKP in database systems. 
%
The interactivity in ZKSQL involves breaking down the computation into smaller sub-circuits to reduce the complexity of the overall circuit. These sub-circuits are then verified interactively, where the prover demonstrates the correctness of each sub-circuit step-by-step. The interaction ensures that the combined outputs of the sub-circuits correspond to the correct execution of the SQL query, allowing the verifier to check the correctness of the entire query process incrementally. 
However, this interactivity increases the communication and computational overhead, as each round of interaction requires multiple exchanges between the prover and verifier.

%
To align with existing research evaluations of ZKP-based databases and to conduct a fair comparison with ZKSQL, we have implemented the six TPC-H queries identified in their evaluation: Q1, Q3, Q5, Q8, Q9, and Q18. 

In addition, we compare our system with Libra~\cite{xie2019libra}, a state-of-the-art non-interactive ZKP system that leverages the GKR protocol~\cite{goldwasser2015delegating}, which is also foundational to vSQL~\cite{zhang2017vsql}. To the best of our knowledge, Libra is the most efficient publicly available system utilizing the GKR protocol. Since the high-level logic for implementing SQL operations in vSQL can be adapted across various ZKP systems, we use Libra’s circuit structure to implement SQL operations based on the logic and optimizations introduced in vSQL. 
%
While Libra uses a fixed input structure, we adopt an alternative approach to avoid the need for relay gates for inputs that may not be needed immediately. That is, we split the circuit into multiple parts, ensuring that each part includes only the necessary inputs in its input layer.

In our experimentation, we adhere to the same query variables wherever applicable, such as the \texttt{orderdate} filter, to maintain consistency and relevance in our results. For query Q9, similar to ZKSQL's approach, we exclude string pattern-matching predicates from our evaluation.  We converted all floating point operations to 64-bit integer ones in our experiments similar to ZKSQL.
%

Our experimental setup quantifies the database scale by the size of the central fact table, \texttt{lineitem}, and scales the dimension tables proportionally, as described in the TPC-H benchmark specifications. We report results across three database sizes---60k Rows, 120k Rows, and 240k Rows---with the \texttt{lineitem} table containing 60k, 120k, and 240k rows, respectively. These varying sizes provide insights into the scalability and practicality of \sysname~ in handling verifiable databases across different volumes of data. Unless otherwise mentioned, our experiments run on 60k Rows.

Our experiments are conducted on Chameleon Cloud~\cite{keahey2020lessons} using a Skylake node, equipped with two Intel Xeon Skylake CPUs running at 2.60 GHz, 192 GB of RAM, and 10 Gigabit Ethernet connectivity.

\vspace{-0.1in}
\subsection{Setup}
\sysname~ eliminates the need for a trusted setup process. Instead, \sysname~ utilizes public parameters. These parameters are essential for both constructing and verifying proofs and are known to all parties involved---the prover and the verifier alike. Importantly, these parameters are not confidential and require no secrecy.

Table~\ref{table:pub_params} details the running time associated with generating these public parameters. It's worth noting that the generation of these parameters is a one-time process; once created, they can be stored and reused indefinitely. The versatility of these parameters allows for their application across various circuits, provided the number of rows in the circuit does not surpass the maximum capacity defined by the public parameters. Consequently, the time spent generating these parameters is not considered part of the cost associated with generating SQL query proofs in this work.

{\bf Running time of database commitment.}  The proof generation for a fixed database commitment can be done once and be reused for SQL queries that are applied on the database. Table~\ref{table:data_commitment}
shows the running time of committing to the $8$ TPC-H tables.

\begin{table}[t]
\captionsetup{skip=2pt} 
\caption {Running time (in seconds) for generating public parameters with different maximal number of rows in Plonkish circuits.}
\label{table:pub_params}
\centering
\renewcommand\arraystretch{1.0}
\begin{tabular}{|c|c|c|c|c|c|}
\hline
\bfseries Maximal number of rows & \bfseries \makecell[c]{$2^{15}$} & \bfseries \makecell[c]{$2^{16}$}  & \bfseries \makecell[c]{$2^{17}$} & \bfseries \makecell[c]{$2^{18}$}  \\\hline

\bfseries \centered{Running time (s)}  & 104  & 221 & 410  & 832  \\\hline

\end{tabular}
\vspace{-0.1in}
\end{table}

\begin{table}[t]
\captionsetup{skip=2pt} 
\caption {Running time (in seconds) of database commitment over data of increasing sizes.}
\label{table:data_commitment}
\centering
\renewcommand\arraystretch{1.0}
\begin{tabular}{|c|c|c|c|c|c|}
\hline
\bfseries The size of the database & \bfseries \makecell[c]{$60k ~Rows$} & \bfseries \makecell[c]{$120k ~Rows$}  & \bfseries \makecell[c]{$240k ~Rows$}   \\\hline

\centered{\bfseries Running time (s) }  & 2.89  & 5.53 & 10.94    \\\hline

\end{tabular}
\vspace{-0.2in}
\end{table}

\subsection{Benchmarking with ZKSQL}
\label{subsec:running_time}

\begin{figure}
\centerline{\psfig{figure=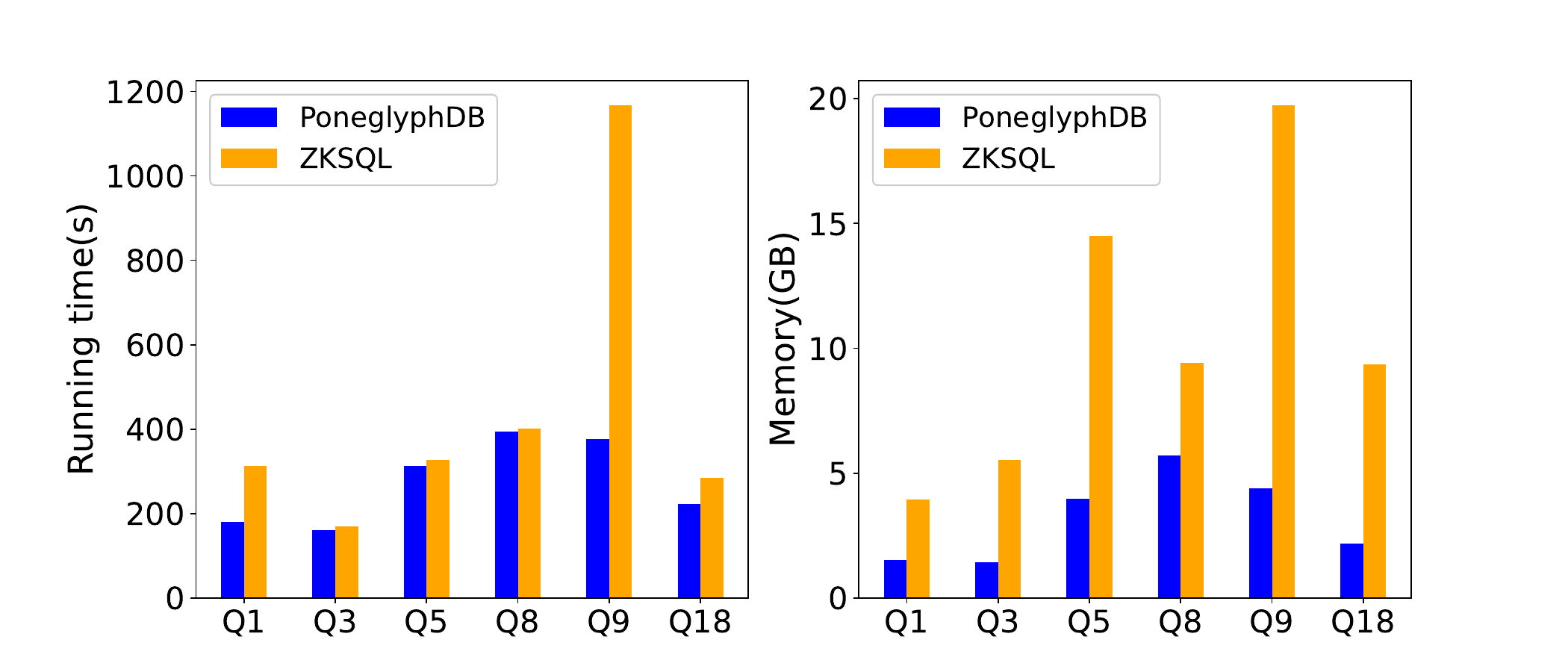, width=0.45\textwidth} }
\captionsetup{skip=2pt} 
\caption{Running time (left figure) and memory usage (right figure) for generating SQL queries proofs.}
\label{fig:combine}
\vspace{-0.2in}
\end{figure}

We compare the running time of generating proofs for six SQL queries in \sysname~ with that of ZKSQL. The results (the left figure of Figure~\ref{fig:combine}), reveal that \sysname---although a non-interactive ZKP solution---achieves performance that is similar to the interactive ZKP solution ZKSQL for most queries. In fact, {\sysname} outperforms ZKSQL significantly---by at least up to 40\%---for queries Q1 and Q9. 
This difference is attributed to the relatively fewer range check (or filtering) and sort operations required in Q1 and Q9. 
%
\sysname~ utilizes arithmetic circuits for handling range checks and sorting operations. Despite the use of lookup tables to optimize the degrees of polynomial constraints and reduce the circuit size, arithmetic circuits can become more complex than boolean circuits, which ZKSQL employs for filtering and sorting operations, especially as the range of data increases. 
Nevertheless, \sysname~ exhibits enhanced performance in join operations, which necessitate arithmetic expressions to represent polynomial constraints accurately. 
Figure~\ref{fig:scale} (right) shows the memory usage for generating proofs for the six SQL queries in \sysname~ and ZKSQL. \sysname~ uses significantly less memory, ranging from $23\%$ to $60\%$ of ZKSQL's usage. 

\begin{figure}
    \centering
    \begin{minipage}[b]{0.45\textwidth}
        \includegraphics[width=\textwidth]{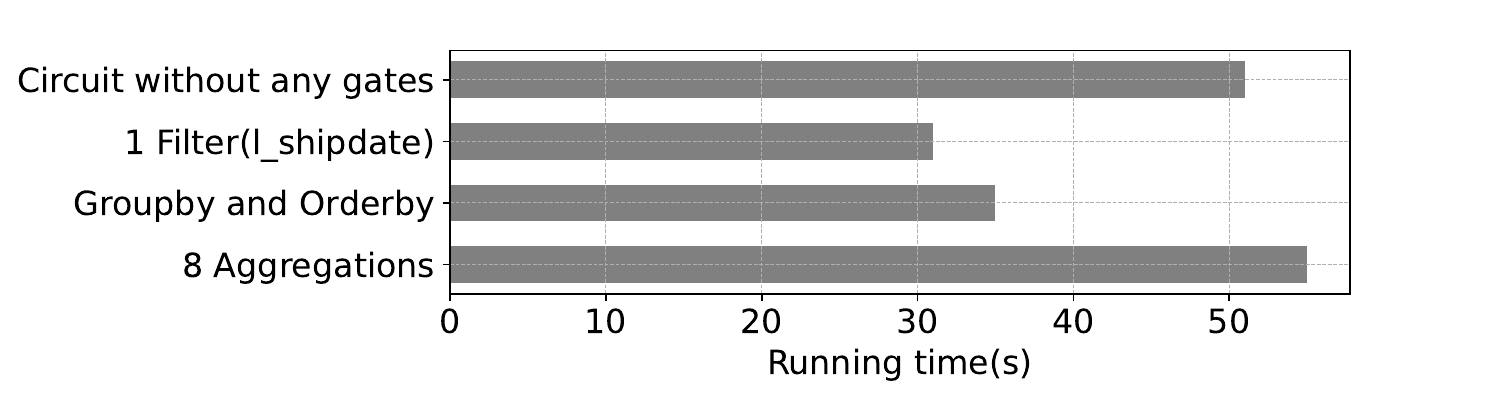}
\vspace{-0.2in}
\captionsetup{skip=5pt} 
	    \caption{{\sysname}'s performance breakdown of different proof generation steps for Q1.}
        \label{fig:q1}
    \end{minipage}
    \hfill 
    \begin{minipage}[b]{0.45\textwidth}
        \includegraphics[width=\textwidth]{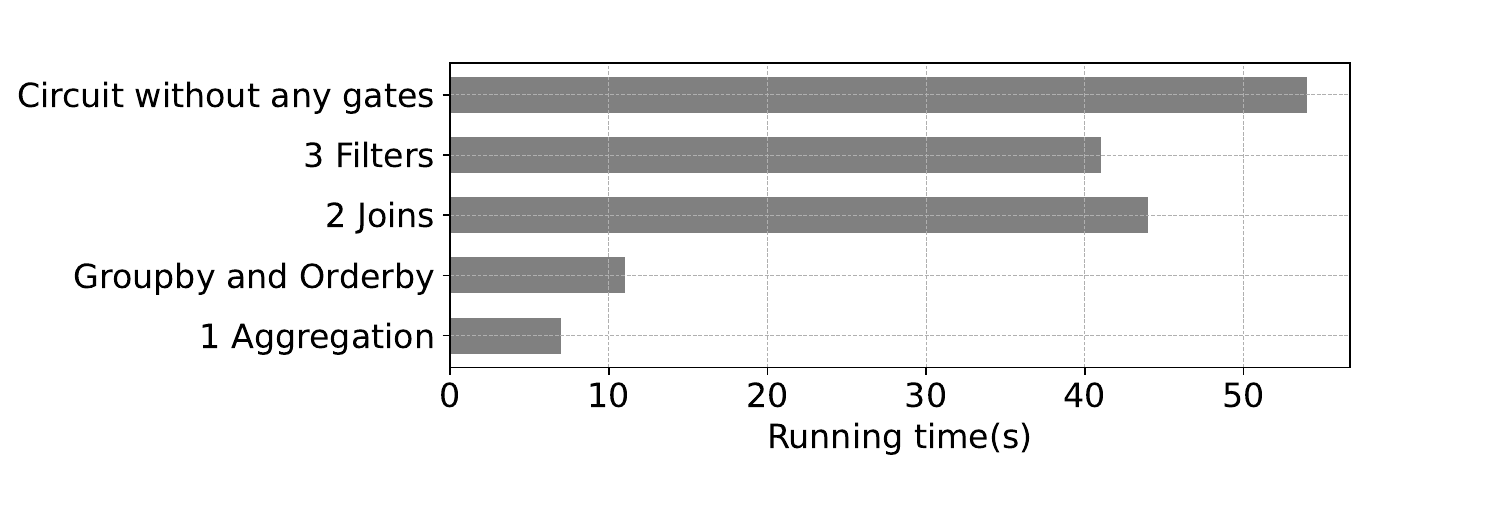}
\vspace{-0.3in}
\captionsetup{skip=5pt} 
	    \caption{{\sysname}'s performance breakdown of different proof generation steps for Q3.}
        \label{fig:q3}
    \end{minipage}
\vspace{-0.2in}
\end{figure}


\subsection{Benchmarking with Libra}
Since Libra is a non-interactive ZKP system, we benchmarked \sysname~ against Libra in terms of three critical factors: proving time, verification time, and proof size. As shown in Table~\ref{table:benchmark}, Libra requires more proving time than \sysname.
In Libra, complex operations such as sorting require a large number of basic gates, with each gate limited to two inputs, which increases both the circuit depth and size. For comparison operations in SQL queries, decimal values are represented using full 64-bit binary representations in Libra. Logical operations on these 64-bit binary numbers necessitate circuits that handle each bit individually, including managing carry bits across the entire bit width. This bitwise processing, along with the overhead of transforming binary values to decimal for subsequent arithmetic operations, results in significantly larger circuits. This increased circuit size leads to longer proving times.
In contrast, \sysname~ optimizes the handling of decimal values by segmenting them into 8-bit chunks and leveraging lookup tables to efficiently validate and perform operations on each segment. The larger circuit size in Libra not only increases the proving time but also leads to longer verification times and larger proof sizes, as shown in Table~\ref{table:benchmark} for queries Q1, Q3, and Q5.

\subsection{Operation Performance}
To enhance our understanding of \sysname's performance, we assess the overheads associated with the steps involved in proof generation. 
%
%
Figures~\ref{fig:q1} and~\ref{fig:q3} show a breakdown of the execution time to generate proofs for queries Q1 and Q3, respectively. We selected these two queries for our performance evaluation because they encompass a comprehensive range of SQL operations, including multiple aggregations, joins, group-by, sort, and filtering functions.

The proof generation process begins with the construction of a comprehensive circuit, encapsulating all facets of witness generation; this preliminary phase is described as a ``circuit without any gates''. Following this, the procedure advances to the integration of polynomial constraints—or gates—that correspond to the SQL operations. These operations include filtering, grouping by, ordering, and performing eight aggregations for query Q1, as well as applying three filters, executing two joins, a group-by, an order-by, and an aggregation for query Q3.

The results show that the initial step takes over 50 seconds, attributable to the fixed overheads determined by the chosen public parameter; a larger public parameter size increases this initial step overhead.
The significant overhead in proof generation, notably from the aggregations in Q1 and the filters and joins in Q3, can be attributed to the extensive computational resources required. Aggregation operations, for instance, necessitate the collation and computation across sizable datasets to yield a singular summary outcome. This task demands multiple iterations of data processing and polynomial constraints verification within the circuit, thereby amplifying its complexity and extending the duration needed for proof generation.
Similarly, filters and joins in Q3 are computationally intensive, as filtering requires checking each record against conditions, while joins involve aligning records from different tables based on join keys.


\begin{table}
\captionsetup{skip=5pt} 
\caption{Benchmarking against Libra in terms of proving time, verification time and proof size.}
\label{table:benchmark}
\setlength{\abovecaptionskip}{0.1cm}
\setlength{\belowcaptionskip}{-0.cm}
\scriptsize
\renewcommand\arraystretch{1.6}
\begin{tabular}{|c|c|>{\centering\arraybackslash}p{1.6cm}|>{\centering\arraybackslash}p{1.6cm}|>{\centering\arraybackslash}p{1.6cm}|}
\hline
\bfseries  & \bfseries  & \bfseries \makecell[c]{Proving time\\(in seconds)}  & \bfseries \makecell[c]{ Verification time\\(in seconds)} & \bfseries \makecell[c]{Proof size\\(in kilobytes)} \\\hline

\multirow{2}{*} {\bfseries Libra} 
& Q1  & 812 & 1.290  & 435.8  \\\cline{2-5}
& Q3  & 997 & 1.212 & 411.4   \\\cline{2-5}
& Q5  & 1021  & 1.227 & 413.9      \\\hline

\multirow{2}{*}{\bfseries \sysname}
& Q1 & 180 & 0.617 & 8.6  \\\cline{2-5}
& Q3 & 161 & 0.725 & 24.7  \\\cline{2-5}
& Q5 & 313 & 0.739 & 29.6     \\\hline
                      
\end{tabular}
\vspace{-0.1in}
\end{table}

\subsection{Scalability}
\begin{figure}
\centerline{\psfig{figure=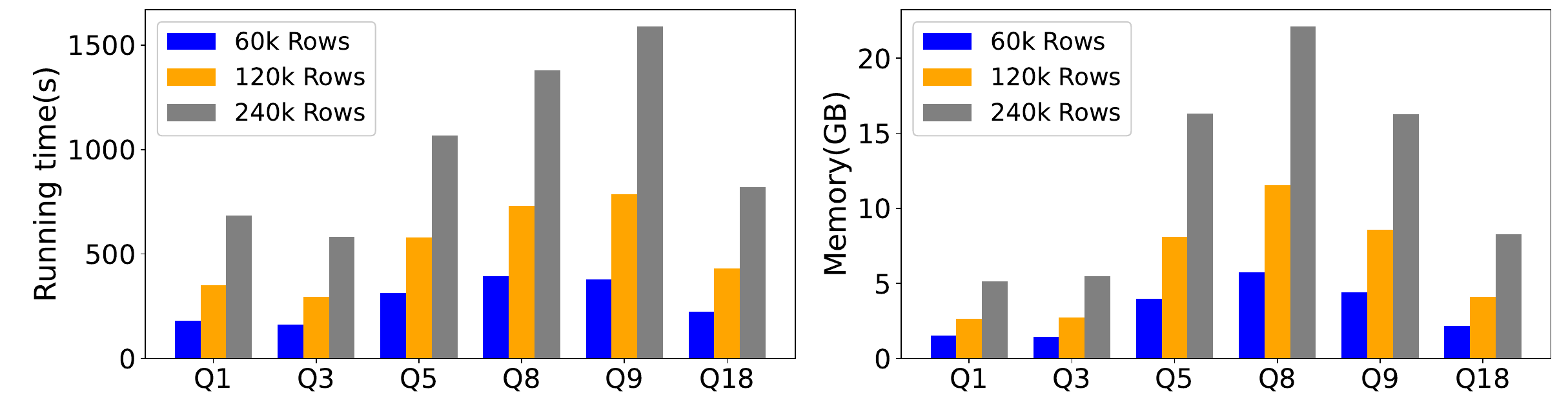, width=0.49\textwidth} }
\captionsetup{skip=5pt} 
\caption{Proof generation time and memory usage over data of increasing sizes.}
\label{fig:scale}
\vspace{-0.2in}
\end{figure}

We evaluate the scalibility of \sysname~ by generating proofs with larger workloads. We evaluate with the workload of TPC-H’s lineitem table at 120k and 240k rows as the lineitem table dominates the complexity of the SQL queries.  
As depicted in Figure~\ref{fig:scale}, the running time and memory consumption increases with the increase in the size of the dataset. Specifically, the running times for the six SQL queries---Q1, Q3, Q5, Q8, Q9, and Q18---exhibit a gradual increase as the database size increases from 60k to 240k Rows. 

The running time for query Q1 starts at 180 seconds for 60k rows and increases to 683 seconds for the 240k row dataset, indicating a proportional relationship between dataset size and observed performance. This is because the size of our circuits linearly grows with the size of inputs and all the polynomial constraints enforced on the circuits have low degrees. This validates our design goal of maintaining low-degree polynomials in {\sysname}'s circuits. 
Memory usage shows a similar pattern, with the memory footprint for query Q1 starting at 1.53 GB for 60k rows and increasing to 5.12 GB for 240k rows. 

\section{Related Work}
\label{sec:related_work}
There is substantial research into verifiable SQL querying, employing a variety of techniques that ensure the integrity and security of query results. These methods can be broadly categorized into three groups: Authenticated Data Structures (ADS)~\cite{tamassia2003authenticated}, Trusted Execution Environments (TEE)~\cite{sabt2015trusted}, and Cryptographic Proof Techniques~\cite{fiege1987zero}.

ADS-based methods use asymmetric cryptography to authenticate data, requiring extra memory to maintain authenticated data structures for SQL query verification \cite{bajaj2013correctdb, yang2009authenticated, zheng2012efficient, papadopoulos2015practical, papadopoulos2014taking}.
While secure, these methods are generally limited to specific computational tasks.


TEE-based approaches, exemplified by \cite{zhou2021veridb, zheng2017opaque, sinha2018veritasdb, bajaj2013correctdb, arasu2017concerto}, secure SQL query results through computations performed within trusted hardware environments.
Basic TEE implementations might expose sensitive data through program traces. Integrating TEE with Oblivious Random Access Machine (ORAM), as in \cite{goodrich2012privacy}, can obscure such traces but at the cost of additional computation time, highlighted in \cite{le2020tale, alam2023sgx}. 


Cryptographic proof techniques, such as zk-SNARKs~\cite{ben2013snarks} and zk-STARKs~\cite{ben2019scalable}, enable entities to verify the correctness of computations without revealing underlying data. 
These techniques have found widespread application across various domains, including blockchain off-chain computations and privacy-preserving machine learning~\cite{zhu2023secure, gu2024zk, lin2024rollstore, gu2023ml}.
These techniques ensure high levels of security but are resource-intensive, requiring substantial memory to manage numerous intermediate values and considerable time to create proofs. 
%

Prior cryptographic proof systems, such as IntegriDB~\cite{zhang2015integridb} and vSQL~\cite{zhang2017vsql}, employ cryptographic verifiable computation to validate a wide range of SQL queries.  While IntegriDB and vSQL ensure data integrity, they operate in an outsourcing model and do not inherently provide zero-knowledge properties. 
An extension of vSQL, referred to as vSQL+~\cite{zhang2017zero}, introduces ZKP; however, it lacks support for ad-hoc queries and does not thoroughly address practical efficiency or the translation of arbitrary SQL statements into cryptographic protocols necessary for such guarantees.
Notably, vSQL and vSQL+ are based on public-coin protocols~\cite{ZK-origins, goldwasser1986private}, which can be transformed into non-interactive ZKP systems using the Fiat-Shamir heuristic~\cite{fiat1986prove}.
%
%
ZKSQL~\cite{li2023zksql} reduces the proving cost by dividing the entire circuit into smaller sub-circuits to reduce the size of the overall circuit. 
%
%
This approach supports ad-hoc queries and maintains zero-knowledge properties, but it shares the common limitations of interactive ZKP. ZKSQL is based on designated-verifier protocols~\cite{goldwasser1986private}, where the Fiat-Shamir heuristic cannot generally be applied to transform the protocol into a non-interactive proof.

Our system, \sysname, generates non-interactive ZKP using recursive proof composition \cite{ben2017scalable, bunz2020recursive, bowe2019recursive, kothapalli2022nova}. It enhances proof generation performance by optimizing arithmetic circuits. 



\section{Conclusion}
\label{sec:conclusion}
We introduce {\sysname}, a non-interactive ZKP-based database designed for efficient confidentiality and provability. {\sysname} optimizes proof generation through recursive methods and tailored designs for SQL queries, showing competitive or superior performance in TPC-H benchmarks compared to existing solutions.


%

\section{Acknowledgments}
This research is partly supported by the NSF under grants CNS1815212 and SaTC-2245372.

\bibliographystyle{ACM-Reference-Format}
\bibliography{paperBib}

\end{document}